\documentclass[%
 reprint,
amsmath,
amssymb,
aps,
longbibliography,
floatfix,
showkeys,
]{revtex4-2}

\usepackage{graphicx}
\usepackage{dcolumn}
\usepackage{bm}
\usepackage{multirow}
\usepackage{textcomp}
\usepackage{gensymb}
\usepackage{lipsum}
\usepackage{longtable}
\usepackage{listings}
\usepackage{booktabs} 
\usepackage{url} 
\usepackage{siunitx} 
\usepackage{graphicx,subfigure,float} 
\usepackage[english]{babel}
\usepackage{gensymb}
\usepackage{fancyhdr}
\usepackage{adjustbox}
\usepackage{mathtools}

\usepackage[T1]{fontenc}
\usepackage[utf8]{inputenc}
\usepackage{amsmath,amssymb,amsfonts,latexsym,cancel}
\usepackage{amsthm} 
\usepackage[bookmarks]{hyperref}
\usepackage{tabularx}
\usepackage{hhline}
\usepackage{times}
\usepackage{graphicx}
\usepackage{color}
\usepackage{wrapfig}
\usepackage{lmodern}

\usepackage{hyperref}

\usepackage{silence}
\begin{document}

\preprint{APS/123-QED}

\makeatletter
\def\l@subsection#1#2{}
\def\l@subsubsection#1#2{}
\def\l@f@section{}%
\def\toc@@font{\footnotesize \sffamily}%
\makeatother

\title{Path Integral for Multiplicative Noise: Generalized Fokker-Planck Equation and Entropy Production Rate in Stochastic Processes With Threshold}

\author{F. S. Abril-Bermúdez}%
\email{fsabrilb@unal.edu.co}%
\email{fab062@inlumine.ual.es}%
\affiliation{%
Department of Economics and Business, Universidad de Almería, Almeria, Spain.\\
Department of Physics, Universidad Nacional de Colombia, Bogota, Colombia.}%

\author{C. J. Quimbay}%
\email{cjquimbayh@unal.edu.co}
\affiliation{%
Department of Physics, Universidad Nacional de Colombia, Bogota, Colombia.}%

\author{J. E. Trinidad-Segovia}%
\email{jetrini@ual.es}
\affiliation{%
Department of Economics and Business, Universidad de Almería, Almeria, Spain.}%

\author{M. A. Sánchez-Granero}%
\email{misanchez@ual.es}
\affiliation{%
Department of Mathematics, Universidad de Almería, Almeria, Spain.}%

\date{\today}

\begin{abstract}
\noindent This paper introduces a comprehensive extension of the path integral formalism to model stochastic processes with arbitrary multiplicative noise. To do so, Itô diffusive process is generalized by incorporating a multiplicative noise term $(\eta(t))$ that affects the diffusive coefficient in the stochastic differential equation. Then, using the Parisi-Sourlas method, we estimate the transition probability between states of a stochastic variable $(X(t))$ based on the cumulant generating function $(\mathcal{K}_{\eta})$ of the noise. A parameter $\gamma\in[0,1]$ is introduced to account for the type of stochastic calculation used and its effect on the Jacobian of the path integral formalism. Next, the Feynman-Kac functional is then employed to derive the Fokker-Planck equation for generalized Itô diffusive processes, addressing issues with higher-order derivatives and ensuring compatibility with known functionals such as Onsager-Machlup and Martin-Siggia-Rose-Janssen-De Dominicis in the white noise case. The general solution for the Fokker-Planck equation is provided when the stochastic drift is proportional to the diffusive coefficient and $\mathcal{K}_{\eta}$ is scale-invariant. Finally, the Brownian motion ($BM$), the geometric Brownian motion ($GBM$), the Levy $\alpha$-stable flight ($LF(\alpha)$), and the geometric Levy $\alpha$-stable flight ($GLF(\alpha)$) are simulated with thresholds, providing analytical comparisons for the probability density, Shannon entropy, and entropy production rate. It is found that restricted $BM$ and restricted $GBM$ exhibit quasi-steady states since the rate of entropy production never vanishes. It is also worth mentioning that in this work the $GLF(\alpha)$ is defined for the first time in the literature and it is shown that its solution is found without the need for Itô's lemma. \vspace{0.25 cm}
\end{abstract}


\maketitle


\section{Introduction}

\noindent A wide variety of systems in physics, finance, and biology are modeled through stochastic processes. Thus, such systems are often influenced by random fluctuations, which can significantly affect their behavior. One of the most powerful mathematical frameworks for analyzing such systems is the path integral formalism, originally developed in quantum mechanics by Feynman \cite{Feynman1948}. This method provides a way to calculate probability amplitudes by summing over all possible paths a system can take between two fixed states $x_{a}$ and $x_{b}$, at times $t_{a}$ and $t_{b}$, respectively \cite{Fai2021}. In the context of the statistical mechanics and stochastic processes, the path integrals estimate the transition probabilities of diffusive systems, leading to a deeper understanding of their probabilistic evolution \cite{Kleinert2002, Kleinert2009, Fai2021}. For instance, by introducing a weight function in the temporal evolution of the moments of the probability distribution for a stochastic process $x(t)$, the origins of the temporal fluctuation scaling and the time evolution of its exponent over time are explained \cite{Abril2021}. Also, it is possible to make the presentation of the path integral from the approach of the supersymmetric theory of stochastic dynamics \cite{Parisi1979, Ovchinnikov2016, Kleinert1997}, which leads to being able to reproduce Kleinert's path integral approach \cite{Kleinert2002, Kleinert2009}, as demonstrated in \cite{Abril2023}. \vspace{0.25 cm}

\noindent When dealing with stochastic differential equations, particularly those driven by noise, an essential distinction arises between \textit{additive} and \textit{multiplicative} noise. In the case of additive noise, the random fluctuations act independently of the state of the system, making the analysis relatively straightforward \cite{Kleinert2002, Abril2023}. However, in many real-world systems, the noise $\eta(t)$ interacts with the system’s dynamics in a state-dependent manner, leading to \textit{multiplicative noise}, such as Itô diffusive processes or a Langevin equation. This type of noise appears in financial markets (e.g., geometric Brownian motion \cite{Giordano2023, Stojkoski2020}), biological systems (e.g., population dynamics \cite{Ricciardi1986}), and anomalous diffusion processes \cite{Quevedo2024, McKinley2018}. Despite its prevalence, the development of path-integral methods for multiplicative noise remains an open challenge. \vspace{0.25 cm}

\noindent Path integrals have been successfully applied in stochastic modeling, notably in birth-death processes, where they enable a Fock space representation that provides insights into non-Markovian effects and continuous field theories \cite{Peliti1985}. Additionally, Weber and Frey have extensively reviewed master equations and their connection to stochastic path integrals, demonstrating how spectral methods, WKB approximations, and variational approaches can be used to solve these equations \cite{Weber2017}. More recently, Del Razo, Lamma, and Merbis introduced a unified framework that merges quantum and field-theoretic approaches to reaction-diffusion systems, offering a consistent representation of multiple stochastic methods \cite{DelRazo2024}. These advancements highlight the versatility of path integrals in stochastic processes and motivate their extension to more general noise structures, as explored in this work. \vspace{0.25 cm}

\noindent A key difficulty in extending the path integral formalism to systems with multiplicative noise lies in its dependence on the interpretation of stochastic calculus. Traditional approaches, such as the Onsager-Machlup (OM) functional \cite{Drr1978, Capitaine1995} and the Martin-Siggia-Rose-Janssen-De Dominicis (MSRJD) formalism \cite{Martin1973, Janssen1976, Janssen1981}, have been successful in describing systems with multiplicative white noise. Nevertheless, these approaches do not fully generalize to arbitrary multiplicative noise processes, as they often rely on specific discretization schemes and truncations that may not preserve all dynamical information \cite{dePirey2022}. Indeed, some studies have shown that the choice of stochastic calculus framework significantly influences the system's statistical properties. For instance, the cross-correlated Gaussian white noises can induce non-equilibrium transitions in stochastic systems, highlighting the importance of considering both Itô and Stratonovich interpretations when modeling multiplicative noise \cite{Denisov2003}. Likewise, the stability analysis of harmonic oscillators influenced by both additive and multiplicative noise reveals that the selected stochastic interpretation plays a crucial role in determining the system's stability \cite{Mendez2014}. \vspace{0.25 cm}

\noindent Furthermore, conventional approaches predominantly focus on systems driven by white noise, often neglecting long-range correlations and colored noise, which are crucial for understanding anomalous diffusion \cite{Kawasaki1973, Quevedo2024}. To address this limitation, the asymptotic behavior of generalized geometric Brownian motion with nonlinear drift has been analyzed, introducing the concept of infinite ergodicity to characterize its long-term dynamics \cite{Giordano2023}. Additionally, a short-time expansion of the Fokker-Planck equation for systems with heterogeneous diffusion has been developed, incorporating arbitrary discretization parameters and spatially dependent noise effects \cite{Dupont2024}. These advancements highlight the necessity for a more comprehensive stochastic path integral formalism that remains independent of specific noise characteristics and discretization choices while ensuring compatibility with well-established theoretical frameworks. \vspace{0.25 cm}

\noindent At once, since the 1960s, with the pioneering work of Mori and Kubo, the Langevin equation has been extended to a generalized Langevin equation (GLE) to account for memory effects and long-range correlations in stochastic systems \cite{Dengler2015}. Over time, the GLE has been applied to a wide range of physical phenomena, including viscoelastic fluids and anomalous diffusion processes \cite{McKinley2018}. This equation is typically formulated as a stochastic integrodifferential equation, where the memory kernel represents the system's response to both external perturbations and internal fluctuations \cite{Dengler2015, McKinley2018}. \vspace{0.25 cm}

\noindent Within this framework, GLEs have been used to model anomalous diffusion, where the mean square displacement of a particle scales sublinearly with time \cite{McKinley2018}. This behavior is a direct consequence of the memory effects encoded in the kernel, which can often be expressed as a power-law or a sum of exponential functions, depending on the system under study \cite{McKinley2018}. The interplay between the mean square displacement and the memory kernel underscores the significance of GLEs in non-Markovian dynamics, where the system's evolution is influenced not only by its present state but also by its entire history \cite{Dengler2015}. Moreover, these GLE models have strong connections to generalized stochastic path integrals, particularly in the presence of nonlinear or multiplicative noise \cite{Dengler2015}. Path integral methods have been employed to derive GLEs in higher-dimensional spaces, incorporating different stochastic computational approaches and extending the Onsager-Machlup formalism \cite{Drr1978, Capitaine1995}. However, these methods are generally restricted to white noise processes, leaving open the challenge of incorporating colored noise or long-range correlations, such as those found in fractional Lévy or Brownian processes \cite{Kawasaki1973, Quevedo2024}. \vspace{0.25 cm}

\noindent An alternative method to solving stochastic processes through path integrals involves the Fokker-Planck equation \cite{Schertzer2004}. The Fokker-Planck equation is a partial differential equation that describes how the transition probability of a particle's velocity evolves under the influence of external and random forces \cite{Fokker1914}. It has been widely applied to model anomalous diffusion, particularly in Lévy $\alpha$-stable flight distributions \cite{Kleinert2002, Schertzer2004, Calvo2009, Ma2018}. Nonetheless, its standard form arises from truncating the Kramers-Moyal expansion, which determines the number of higher-order spatial derivatives included in the equation \cite{Kramers1940, Moyal1949}. This truncation is done until the second order, following the Pawula theorem, which states that if any higher-order term is nonzero, all must be present \cite{Risken1996}. For most continuous Markov processes, this truncation is justified, leading to the standard diffusion equation \cite{Fokker1914, Schertzer2004}. In contrast, systems exhibiting non-Gaussian noise, jump processes, or long-range correlations require higher-order terms, resulting in generalized Fokker-Planck equations that account for memory effects and discontinuous dynamics. These extensions are particularly relevant in anomalous diffusion and non-equilibrium statistical mechanics \cite{Wio2012, Laskin2002}. \vspace{0.25 cm}

\noindent On the other hand, to address non-Markovian processes within path integral formulations, several approaches have been proposed \cite{Wio2012, Jung1987, Hanggi2007, Tsallis1988, Curado1991, Curado1992, Borland1998, GellMann2004}. Among these, the Unified Colored Noise Approximation is one of the most widely adopted methods \cite{Wio2012, Jung1987, Hanggi2007}. This technique expands the stochastic system’s state space by introducing additional random variables, allowing for the integration of functions over newly defined stochastic variables. However, applying adiabatic elimination to simplify the problem often removes higher-order temporal derivatives and terms involving the first derivative, leading to a Markovian approximation of the original non-Markovian process. \vspace{0.25 cm}

\noindent An alternative strategy for handling long-range temporal correlations involves using fractional operators \cite{Wio2012, Calvo2009, Ma2018, Eab2006, Bhrawy2017}, particularly in the study of fractional stable processes and fractional Lévy motion. Nevertheless, the $\alpha$-stable Lévy distributions, commonly employed in such models, exhibit diverging variances and moments of higher order \cite{Piryatinska2005, Schertzer2004}. Hence, alternative probability distributions with finite moments at all orders, such as truncated or tempered Lévy distributions, should be explored. \vspace{0.25 cm}

\noindent A notable and well-known application of stochastic processes within this context is the Feynman-Kac formula, which establishes an equivalence between parabolic partial differential equations and stochastic processes under the framework of the Feynman path integral \cite{Kac1949, Wu2016, Capuozzo2021}. Specifically, it has been shown that the Feynman-Kac formula is equivalent to the Fokker-Planck equation associated with the process, except that it is formulated with final conditions rather than initial conditions \cite{Kac1949}. Additionally, this formulation has been extended to symmetric Lévy flights and, consequently, to fractional operators \cite{Tarasov2008, Aljethi2023}. Even so, the broader question of whether this generalization can be made independent of the specific distribution type and derivative order remains unresolved. \vspace{0.25 cm}

\noindent Thence, this work aims to develop a generalized path integral formalism for stochastic systems driven by arbitrary multiplicative noise, independent of the chosen stochastic calculus parameterization. Our approach builds upon the supersymmetric theory of stochastic dynamics \cite{Parisi1979, Ovchinnikov2016}, which provides a unifying framework for different interpretations of stochastic processes. Using the Parisi-Sourlas method, we construct a multiplicative stochastic path integral that generalizes the Fokker-Planck equation for Itô diffusive processes without requiring truncation at a specific derivative order. This approach ensures consistency with known results like the OM and MSRJD functionals, and the Fokker-Planck equation for Itô diffusive processes while extending their applicability to a broader class of stochastic systems, enhancing the theoretical foundation for modeling complex noise-driven dynamics. \vspace{0.25 cm}

\noindent A fundamental aspect of our formulation is the introduction of a noise cumulant generating function, which accounts for higher-order noise statistics and enables a more accurate description of non-stationary time series \cite{Kleinert2002, Kleinert2009, Abril2021}. We also introduce a discretization parameter $\gamma\in[0,1]$ to interpolate between different stochastic calculus interpretations, such as Itô integral ($\gamma=0$) \cite{Ito1950}, Fisk-Stratonovich integral ($\gamma=0.5$) \cite{Stratonovich1966}, and Hänggi-Klimontovich integral ($\gamma=1$) \cite{Hanggi1982}. This flexibility ensures that our formalism remains compatible with various modeling approaches while providing a more general and physically consistent description of stochastic dynamics. \vspace{0.25 cm}

\noindent To illustrate the effectiveness of our method, we apply it to several well-known stochastic processes with threshold conditions, including Brownian motion (BM), geometric Brownian motion (GBM), and Lévy $\alpha$-stable flights (LF($\alpha$)). Additionally, we introduce and analyze a new stochastic process, geometric Lévy $\alpha$-stable flight (GLF($\alpha$)), which generalizes GBM in the presence of heavy-tailed distributions. By deriving exact solutions for these processes, we compute key statistical properties such as probability densities, Shannon entropy, and entropy production rates, highlighting the impact of multiplicative noise on their long-term behavior. \vspace{0.25 cm}

\noindent Furthermore, our analysis reveals an important connection between entropy production and non-equilibrium steady states in stochastic systems. In traditional thermodynamics, the principle of minimum entropy production proposed by Prigogine, suggests that when a system reaches equilibrium, the rate at which entropy is produced decreases to a minimum value, corresponding to a balance of forces in the system where no more macroscopic changes occur \cite{Klein1954, Tom2012, Spinney2012}. Thus, a state of minimum entropy production also corresponds to a state of maximum entropy which is the one that best represents the current state of knowledge about a system (equilibrium steady state) \cite{Jaynes1957, Jaynes1957_2}. Nonetheless, the entropy production does not necessarily vanish at a steady state (non-equilibrium steady state) for a multiplicative stochastic process. Instead, we demonstrate that a finite entropy production rate persists in certain cases, reflecting ongoing fluctuations and energy exchange within the system. \vspace{0.25 cm}

\noindent Indeed, for systems modeled by Langevin equations or described via generalized stochastic dynamics, the rate of entropy production serves as an indicator of the distance to equilibrium. In these cases, a nonzero entropy production rate signifies ongoing irreversible processes or dissipation, while a zero rate indicates a steady state, the equilibrium steady state \cite{Tom2012, Spinney2012}. For example, in systems governed by the Fokker–Planck equation, the rate of entropy production decreases as the system approaches its steady state, where fluctuations are balanced by dissipation \cite{Tom2012}. Indeed, in systems governed by stochastic differential equations with multiplicative noise, total entropy production is decomposed into three distinct components. Two of these components adhere to fluctuation theorems, ensuring they remain positive on average, as demonstrated through integral fluctuation relations. The third component, however, cannot be attributed solely to irreversibility or relaxation processes, reflecting more complex dynamics that go beyond traditional interpretations of entropy production in non-equilibrium systems \cite{Spinney2012}. Thence, this relationship is also important for understanding anomalous diffusion processes, where non-Gaussian behaviors or long-range correlations can influence how quickly or slowly a system approaches its steady state. \vspace{0.25 cm}

\noindent The structure of this paper is as follows: Section \ref{Stochastic processes} introduces the generalized Itô diffusive process and the cumulant generating function. Section \ref{Feynman Path Integral} presents the construction of the multiplicative stochastic path integral, accounting for various parameterizations of the stochastic calculus. Section \ref{Drift-free Stochastic path integral} explores a specific class of systems exhibiting scale-invariant behavior and solves the path integral explicitly without the need of Itô's lemma. Section \ref{FP-GIDP} derives the generalized Fokker-Planck equation, demonstrating its relationship with the Feynman-Kac formula and its dependence on stochastic calculus interpretations. Finally, Section \ref{Stochastic Processes application} applies our formalism to the four types of stochastic processes with thresholds, comparing the numerical results to analytical solutions for probability density functions, Shannon entropy, and entropy production rate. \vspace{0.07 cm}

\section{\label{Stochastic processes}Stochastic processes}

\noindent A stochastic process in $\mathbb{R}$ is a collection of random variables $\left\{X_{t}:\Omega\rightarrow\mathbb{R}\right\}_{t\in\mathcal{S}}$ defined over the same probability space $(\Omega, \mathcal{F}, \mathbb{P})$, where $\Omega$ is the sample space, $\mathcal{F}$ is a $\sigma$-algebra over $\Omega$, and $\mathbb{P}$ is a probability measure over $\mathcal{F}$. Thus, it is common for the set of indexes ($\mathcal{S}$) to be taken as time in many practical applications such that $\mathcal{S}=\left[t_{0}, T\right]$, with $T>t_{0}\geq0$. It is also important to mention that formally, for each $t\in\mathcal{S}$, $X_{t}$ is a function $X_{t}: \Omega\to\mathbb{R}$ that assigns a real value to each element of the sample space $\Omega$, and $X_{t}(\omega)$ is called the realization of the stochastic process over $\omega\in\Omega$. \vspace{0.25 cm}

\noindent In addition to its basic definition, many stochastic processes are also defined as the solution of a stochastic differential equation (SDE). Indeed, a well-known case is the Itô diffusive processes that satisfy the SDE:
\begin{equation}
\label{Eq. Cumulant Generating Function 1}
dX_{t}=\mu\left(X_{t},t\right)dt+\sigma\left(X_{t},t\right)dW_{t},
\end{equation}
\noindent where $\mu(\cdot,t)$ and $\sigma(\cdot,t)$ are given functions, and $W_t$ represents a Wiener process or Brownian motion. These equations extend the concept of ordinary differential equations to include terms that model the randomness inherent in many physical phenomena, such as the generalized Langevin equation \cite{Dengler2015, McKinley2018, Kawasaki1973, Quevedo2024}; biological phenomena, such as the Volterra equations and population dynamics \cite{Ricciardi1986}; and economic phenomena \cite{Friedrich2011}, such as the Feynman-Kac formula \cite{Kac1949, Wu2016, Capuozzo2021, Tarasov2008, Aljethi2023}, among others. \vspace{0.25 cm}

\noindent Thus, an Itô diffusive process can be extended by the following SDE:
\begin{equation}
\label{Eq. Cumulant Generating Function 2}
\frac{dX(t)}{dt}=\mu\left[X(t),t\right]+\sigma\left[X(t),t\right]\eta(t),
\end{equation}
\noindent where $\eta(t)$ represents an arbitrary noise defined only by its statistical properties, specifically by its cumulant generating function, as seen later. Note that Eq. \eqref{Eq. Cumulant Generating Function 1} is a special case of Eq. \eqref{Eq. Cumulant Generating Function 2} when $\eta(t)=\frac{dW_{t}}{dt}$, i.e., the white noise case. Thus, it is worth noting that from a mathematical point of view, the Wiener integral is a well-defined object (see $dW_{t}$ in Eq. \eqref{Eq. Cumulant Generating Function 1}), while white noise is not always well-defined since in the sense of usual calculus $\frac{dW_{t}}{dt}=\lim_{\tau\to0}{\frac{W(t+\tau)-W(t)}{\tau}}=\mathcal{O}\left(\tau^{-1/2}\right)$ does not exist. \vspace{0.25 cm}

\noindent Therefore, it is worth mentioning that without abusing the notation, it is understood that $dX(t)$ is the differential increment between the random variables $X(t)$ and $X(t+dt)$, with $dt>0$, even when the quotient $\frac{dX(t)}{dt}$ is not always correctly defined in the sense of the usual calculus. From now on, Eq. \eqref{Eq. Cumulant Generating Function 2} is called a generalized Itô diffusive process (GIDP) and can accommodate cases like a generalized Langevin equation \cite{Parisi1979, Ovchinnikov2016, Kleinert1997}. \vspace{0.25 cm}

\noindent Now, the probability density function of the random variable $z\equiv\eta_{t}=\eta(t)$ is defined as \cite{Blanco_Castaneda2012, Abril2021}
\begin{equation}
\label{Eq. Cumulant Generating Function 3}
D_{\eta}(z) = \int_{\mathbb{R}} \frac{dp}{2\pi} e^{ipz} \widetilde{D}_{\eta}(p) = \int_{\mathbb{R}} \frac{dp}{2\pi} e^{ipz} e^{\mathcal{K}_{\eta}(p)},
\end{equation}
\noindent where $\eta$ represents a known underlying noise in the system (see Eq. \eqref{Eq. Cumulant Generating Function 2}), $\widetilde{D}_{\eta}(p)$ is the characteristic function, corresponding to the Fourier components of $D_{\eta}(z)$. The function $\mathcal{K}_{\eta}(p)$ defines the cumulant generating function (CGF) \cite{Stuart1994}. \vspace{0.25 cm}

\noindent Consequently, the moments about the origin of the distribution are derived from the CGF $\mathcal{K}_{\eta}(p)$. For all $n\in\mathbb{N}$, these moments are given by
\begin{equation}
\label{Eq. Cumulant Generating Function 4}
\mathbb{E}\left[z^{n}\right] = \int_{\mathbb{R}} z^{n}D_{\eta}(z)dz=\left.(-i)^{n}\frac{d^{n}}{dp^{n}}e^{\mathcal{K}_{\eta}(p)}\right|_{p=0}.
\end{equation}
\noindent Then, without loss of generality, it is assumed that the CGF admits an analytical representation of the form
\begin{equation}
\label{Eq. Cumulant Generating Function 5}
\mathcal{K}_{\eta}(p) = \ln{\left[\widetilde{D}_{\eta}(p)\right]}=\sum_{n=1}^{\infty}c_{n}\frac{(ip)^{n}}{n!},
\end{equation}
\noindent where $c_{n}= (-i)^{n}\mathcal{K}_{\eta}^{(n)}(0)$ are the cumulants of the probability density function, corresponding to the coefficients in the power series of the CGF. Furthermore, it is important to define the normalization condition of the probability density function given by the expression
\begin{equation}
\label{Eq. Cumulant Generating Function 6}
\int_{-\infty}^{\infty}D_{\eta}(z)dz=1.
\end{equation}
\noindent The above implies that $\mathcal{K}_{\eta}(0)=c_{0}=0$ since $e^{\mathcal{K}_{\eta}(0)}=\widetilde{D}_{\eta}(0) = \int e^{-i0z} D_{\eta}(z)dz = 1$. \vspace{0.07 cm}

\section{\label{Feynman Path Integral}Feynman path integral}

\noindent This section outlines the construction of the stochastic path integral for a GIDP. To establish a solid foundation, we first revisit the fundamental concepts of the Feynman path integral, originally formulated in terms of positions, momenta, and the Hamiltonian function. In the following section \ref{Stochastic path integral}, the path integral for the framework of stochastic processes is introduced, emphasizing its key similarities and differences with the Feynman path integral, which we will refer to as the standard or usual path integral throughout this work. \vspace{0.25 cm}

\noindent In the functional formalism, Feynman's kernel or amplitude of transition probability from a position $q_{a}$ in the time $t_{a}$ to a position $q_{b}$ in the time $t_{b}$, is defined as \cite{Feynman1948}
\begin{equation}
    \label{Eq. Feynman path integral definition 1}
    \mathcal{A}\left(q_{b},t_{b};q_{a},t_{a}\right)= \int_{q_{a}}^{q_{b}}\mathcal{D}q \int\frac{\mathcal{D}p}{2\pi}e^{-\frac{i}{\hbar}\mathcal{S}\left[p(t),q(t)\right]},
\end{equation}
\noindent where $\mathcal{D}q$ and $\mathcal{D}p$ correspond to functional measures in the space of positions and momenta, respectively \cite{Courant1989, Popov2001, Daniell1919}, and $\mathcal{S}\left[p(t),q(t)\right]=\int_{t_{a}}^{t_{b}}\left[p\dot{q}-H\left[p(t),q(t)\right]\right]dt$ correspond to the classical action of the system \cite{Zia2009}. The transition amplitude arises from the interference of possible paths in the configuration space with fixed endpoints. Each path's contribution is proportional to the exponential of its classical action. While all possible paths contribute to the integral, those closer to the classical path have a more significant impact \cite{Feynman1948, Fai2021}. In the imaginary time formalism, the Feynman kernel is \cite{Fai2021}
\begin{equation}
    \label{Eq. Feynman path integral definition 2}
    \begin{split}
    \mathcal{A}\left(q_{b},\tau_{b};q_{a},0\right)&= \int\mathcal{D}q\int\frac{\mathcal{D}p}{2\pi}e^{-\frac{1}{\hbar}\int_{0}^{\tau_{b}}d\tau\left[ip
    \dot{q}-H\left[p,q\right]\right]},
    \end{split}
\end{equation}
\noindent where $q=q(\tau)$, $p=p(\tau)$, $t=t_{a}-i\tau$, and $\tau_{b}=i(t_{b}-t_{a})$. We note that Eq. \eqref{Eq. Feynman path integral definition 2} has a structure similar to an inverse Fourier transform. \vspace{0.25 cm}

\noindent Also, the Wick rotation allows the partition function of a quantum system, $\mathcal{Z}=\text{Tr}\left[e^{-\beta\hat{H}}\right]$, to be expressed using path integrals by relating it to the quantum mechanical evolution operator $e^{-i(t-t_{a})\hat{H}}$. Replacing real time with imaginary time leads to the correspondence $\tau=\hbar\beta$, where $\beta^{-1}=k_{B}T$ represents the thermal energy at temperature $T$ \cite{Fai2021}. It is worth emphasizing that the partition function links microscopic states to macroscopic observables by summing over all system configurations weighted by Boltzmann factors such as the usual path integral. It also normalizes the probability density function, ensuring proper statistical weighting of states. \vspace{0.07 cm}

\subsection{\label{Stochastic path integral}Extension of the path integral for a generalized Itô diffusive process}

\noindent The extension of the path integral formalism to stochastic processes has been addressed through the supersymmetric theory of stochastic dynamics \cite{Kleinert2002, Abril2023}. This extension specifically applies to the evolution of additive stochastic variables, where $\sigma(\cdot,t)=1$ in the expression \eqref{Eq. Cumulant Generating Function 2}. Conversely, when $\sigma\left(\cdot,t\right)\neq1$, the noise is considered multiplicative, and $\sigma\left(\cdot,t\right)$ is referred to as the diffusive coefficient of the system. Moreover, this path integral approach reveals that $\mu\left(\cdot,t\right)$, the stochastic drift, is proportional to the initial average value of the data, i.e., $\mathbb{E}\left[X\left(t\right)\right]\sim \mu\left[X(t),t\right]$. \vspace{0.25 cm}

\noindent The supersymmetric theory of stochastic dynamics offers a rigorous framework for analyzing continuous-time stochastic differential equations, both in the presence and absence of noise \cite{Ovchinnikov2016}. This approach is motivated by the observation that many natural systems exhibit long-range correlations, which can be interpreted in terms of gapless excitations such as Nambu-Goldstone particles \cite{Ovchinnikov2016}. According to Goldstone's theorem, the spontaneous breaking of a continuous symmetry leads to massless or nearly massless scalar excitations \cite{Goldstone1961}. As a result, long-range correlations in stochastic systems can be understood as arising from the spontaneous breaking of symmetry or supersymmetry, a phenomenon observed in certain stochastic differential equations \cite{Ovchinnikov2016}. \vspace{0.25 cm}

\noindent The study of supersymmetry in stochastic differential equations originates from the Parisi-Sourlas stochastic quantization procedure \cite{Parisi1979, Ovchinnikov2016, Kleinert1997}. This approach demonstrates that a classical field theory with random sources is perturbatively equivalent to the corresponding quantum field theory in two fewer dimensions \cite{Parisi1979}. More importantly, it establishes a direct connection between the functional formalism of the path integral and Langevin equations, a key aspect exploited in this work \cite{Kleinert1997}. Building on this foundation, the construction of the stochastic path integral for a GIDP involves considering all possible trajectories for the stochastic variable $X(t)$, ensuring that it reaches the value $X_{f}$ at time $t_{f}$ given that it had a value $X_{0}$ at an initial time $t_{0}$, where $0\leq t_{0}<t_{f}\leq T$. Similar to the standard path integral, which sums over all possible trajectories in phase space with fixed positions $q_{a}$ and $q_{b}$ (see Eq. \eqref{Eq. Feynman path integral definition 1}), the stochastic path integral for a GIDP must account for all realizations of $X(t)$ that satisfy the boundary conditions $X_{0}$ and $X_{f}$. However, to properly capture the system fluctuations, the noise term $\eta(t)$ is averaged, ensuring that only trajectories consistent with the SDE \eqref{Eq. Cumulant Generating Function 2} are considered. \vspace{0.25 cm}

\noindent Therefore, the gauge condition of the generalized Itô diffusive process is such that the transition probability for the stochastic variable $X(t)$ to acquire a value $X_{f}$ at time $t_{f}$ given that it had a value $X_{0}$ at time $t_{0}$, with $0\leq t_{0}<t_{f}\leq T$, is given by
\begin{widetext}
\begin{align}
    \label{Eq. STS path integral 1}
    \mathcal{P}_{\eta}(X_{f},t_{f}|X_{0},t_{0})&=\left\langle\int_{X(t_{0})=X_{0}}^{X(t_{f})=X_{f}}\mathcal{J}\left[X(t),\eta(t)\right]\delta\left[\frac{dX(t)}{dt}-\mu\left[X(t),t\right]-\sigma\left[X(t),t\right]\eta(t)\right]\mathcal{D}X\right\rangle_{\eta(t)},\\
    \label{Eq. STS path integral 2}
    &=\int_{X(t_{0})=X_{0}}^{X(t_{f})=X_{f}} \int\mathcal{D}\eta\;\mathbb{P}\left[\eta(t)\right]\mathcal{J}\left[X(t),\eta(t)\right] \delta\left[\frac{dX(t)}{dt}-\mu\left[X(t),t\right]-\sigma\left[X(t),t\right]\eta(t)\right]\mathcal{D}X.
\end{align}
\end{widetext}
\noindent where $\mathbb{P}\left[\eta(t)\right]$ is the normalized distribution of noise patterns, $\left\langle\cdots\right\rangle_{\eta(t)}$ is the averaging over noise $\eta(t)$, and $\mathcal{J}\left[X(t),\eta(t)\right]$ is the determinant of the Jacobian between variables $X(t)$ and $\eta(t)$. Thus, $\mathcal{J}\left[X(t),\eta(t)\right]$ allows that Eq. \eqref{Eq. Cumulant Generating Function 2} is inverted so short-time propagators are expressed only in terms of $X(t)$. Similarly, the measure $\mathcal{D}\rho=\mathbb{P}\left[\eta(t)\right]\mathcal{D}\eta$, known as the measure of the process \eqref{Eq. STS path integral 1}, assigns the probability of occurrence for each trajectory in the phase space defined by $X(t)$ and $\eta(t)$. Finally, all such trajectories are filtered using the Dirac delta functional to ensure they satisfy the evolution rule of a GIDP (see Eq. \eqref{Eq. Cumulant Generating Function 2}). \vspace{0.25 cm}

\noindent Now, when the extension of Eq. \eqref{Eq. Cumulant Generating Function 3} to the continuum is made, the CGF satisfies the following identity
\begin{equation}
    \label{Eq. STS path integral 3}
    e^{\int_{t_{0}}^{t_{f}} \mathcal{K}_{\eta}\left[p(t)\right]dt}=\int\mathcal{D}\eta\;e^{-i\int_{t_{0}}^{t_{f}} p(t)\eta(t)dt}\mathbb{P}\left[\eta(t)\right].
\end{equation}
\noindent It is important to highlight that Eq. \eqref{Eq. STS path integral 3} is valid for any almost everywhere continuous function $p(t)$. Also, since the Dirac delta functional satisfies the following expression,
\begin{equation}
    \label{Eq. STS path integral 4}
    \delta\left[f(t)\right]=\int\frac{\mathcal{D}p}{2\pi}\;e^{i\int_{t_{0}}^{t_{f}} p(t)f(t)dt},
\end{equation}
\noindent for all continuous function $f$ on $\left[t_{0},t_{f}\right]$, we have that the transition probability of the stochastic variable $X(t)$ becomes
\begin{widetext}
\begin{align}
    \mathcal{P}_{\eta}(X_{f},t_{f}|X_{0},t_{0})&=\int_{X(t_{0})=X_{0}}^{X(t_{f})=X_{f}} \int\mathcal{D}\eta\;\mathbb{P}\left[\eta\right]\mathcal{J}\left[X,\eta\right]\int\frac{\mathcal{D}p}{2\pi}e^{i\int_{t_{0}}^{t_{f}} p(t)\left[\frac{dX(t)}{dt}-\mu\left[X(t),t\right]-\sigma\left[X(t),t\right]\eta(t)\right]dt}\mathcal{D}X\notag\\
    &= \int_{X_{0}}^{X_{f}}\mathcal{D}X\int\frac{\mathcal{D}p}{2\pi}e^{\int_{t_{0}}^{t_{f}}\left[ip(t)\frac{dX(t)}{dt}-ip\mu\left[X(t),t\right]\right]dt}\int\mathcal{D}\eta \;\mathcal{J}\left[X(t),\eta(t)\right]e^{-i\int_{t_{0}}^{t_{f}}p(t)\sigma\left[X(t),t\right]\eta(t)dt}\mathbb{P}\left[\eta(t)\right]\notag\\
    &= \int_{X_{0}}^{X_{f}}\mathcal{D}X\int\frac{\mathcal{D}p}{2\pi}e^{\int_{t_{0}}^{t_{f}}\left[ip(t)\frac{dX(t)}{dt}-ip(t)\mu\left[X(t),t\right]\right]dt}\mathcal{J}\left[X(t)\right]e^{\int_{t_{0}}^{t_{f}}\mathcal{K}_{\eta}\left[\sigma\left[X(t),t\right]p(t)\right]dt}\notag\\
    \label{Eq. STS path integral 5}    &=\int_{X_{0}}^{X_{f}}\mathcal{D}X\int\frac{\mathcal{D}p}{2\pi}\mathcal{J}\left[X\right]\exp{\left\{\int_{t_{0}}^{t_{f}}\left[ip(t)\frac{dX(t)}{dt}-ip(t)\mu\left[X(t),t\right]+\mathcal{K}_{\eta}\left[\sigma\left[X(t),t\right]p(t)\right]\right]dt\right\}}.
\end{align}
\end{widetext}
\noindent Hence, Eq. \eqref{Eq. STS path integral 5} indicates that the transition probability amplitude from configuration $X_{0}$ at time $t_{0}$ to configuration $X_{f}$ at time $t_{f}$ is largely determined by the CGF of the underlying noise $\eta(t)$ in the system. In the case of multiplicative noise, it is particularly emphasized that the CGF argument is explicitly evaluated in the diffusion coefficient $\sigma\left[X(t),t\right]$. Then, replacing the value of the $\mathcal{J}\left[X\right]$ (see Appendix \ref{Appendix A}), as an additional phase entering the exponential of the path integral gives that the stochastic path integral for a GIDP is
\begin{equation}
    \label{Eq. STS path integral 6}
    \mathcal{P}_{\eta}\left(X_{f},t_{f}|X_{0},t_{0}\right) =\int_{X_{0}}^{X_{f}}\mathcal{D}X\int\frac{\mathcal{D}p}{2\pi}e^{\mathcal{S}_{\eta}\left[\dot{X}(t),X(t)\right]},
\end{equation}
\noindent where $\dot{X}(t)=\frac{dX(t)}{dt}$, and
\begin{equation*}
    \mathcal{S}_{\eta}\left[\dot{X}(t),X(t)\right]=\int_{t_{0}}^{t_{f}}\mathcal{L}_{\eta}\left[\dot{X}(t),X(t)\right]dt,
\end{equation*}
\noindent is the action defined by the Lagrangian
\begin{align}
    \label{Eq. STS path integral 7}
    \mathcal{L}_{\eta}&= ip\left(\dot{X}-\mu\right)+\mathcal{K}_{\eta}\left[\sigma p\right]-\gamma\sigma\frac{\partial}{\partial X}\left(\frac{\mu}{\sigma}\right).
\end{align}
\noindent It is important to highlight that $\gamma\in\left[0,1\right]$ represents a parameterization of stochastic calculus such that $\gamma=0$ corresponds to the Itô calculus \cite{Ito1950}, $\gamma=1/2$ corresponds to the Fisk-Stratonovich calculus \cite{Stratonovich1966}, and $\gamma=1$ corresponds to the Hänggi-Klimontovich calculus \cite{Hanggi1982}. \vspace{0.25 cm}

\noindent Thus, the stochastic path integral for a GIDP has the same structure as a standard path integral, i.e., both are a superposition of trajectories with fixed endpoints in a phase space where each trajectory has a weight proportional to the classical action of the system. Also, the stochastic path integral for a GIDP resembles the path integral of statistical mechanics in that it estimates averages over noise realizations proportional to the probability density function of the system (see Eq. \eqref{Eq. Feynman path integral definition 2}). Still, the two main differences are: \vspace{0.25 cm}

\begin{enumerate}
    \item {The usual path integral considers continuous and smooth paths, while the stochastic path integral for a GIDP considers continuous and often non-differentiable paths. \vspace{0.25 cm}
    }
    \item {The usual path integral involves the integration of complex exponentials of the classical action, which leads to quantum interference effects, while the stochastic path integral for a GIDP usually involves real-valued functionals that depend on the CGF of the noise $\eta(t)$ and the type of stochastic calculation used measured trough $\gamma\in[0,1]$. \vspace{0.25 cm}
    }
\end{enumerate}

\noindent Furthermore, if $\mathcal{L}_{\eta}$ is considered as a usual Lagrangian, the canonically conjugated moment corresponds to
\begin{equation}
    \label{Eq. STS path integral 8}
    p_{X}=\frac{\partial\mathcal{L}_{\eta}}{\partial\dot{X}}=ip,
\end{equation}
\noindent which implies a Hamiltonian of the form
\begin{align}
    \mathcal{H}_{\eta}&=p_{X}\dot{X}-\mathcal{L}_{\eta}\notag\\
    \label{Eq. STS path integral 9}
    &=p_{X}\mu-\mathcal{K}_{\eta}\left[-i\sigma p_{X}\right]+\gamma\sigma\frac{\partial}{\partial X}\left(\frac{\mu}{\sigma}\right).
\end{align}
\noindent Therefore, it is important to highlight the results of Eq. \eqref{Eq. STS path integral 9} and \eqref{Eq. STS path integral 7} in terms of other functionals known in stochastic processes such as the Onsager-Machlup (OM) \cite{Drr1978, Capitaine1995}, and the Martin-Siggia-Rose-Janssen-De Dominicis (MSRJD) action \cite{Martin1973, Janssen1976, Janssen1981}. In fact, if $\eta(t)$ represents a white noise with diffusion coefficient $\sigma_{0}\left(X\right)$ and CGF $\mathcal{K}_{\eta}\left(p\right)=-\frac{1}{2}p^{2}$, it follows that the Lagrangian $\mathcal{L}_{\eta}$, and the Hamiltonian $\mathcal{H}_{\eta}$, satisfy the expressions
\begin{align}
    \label{Eq. STS path integral 10}
    \mathcal{L}_{\eta}\left[\dot{X},X\right]&= \frac{\left(\dot{X}-\mu\right)^{2}}{2\sigma_{0}^{2}}-\gamma\frac{\partial\mu}{\partial X}+\gamma\mu\frac{\partial\ln{\sigma_{0}}}{\partial X}\\
    \label{Eq. STS path integral 11}
    \mathcal{H}_{\eta}\left[p_{X},X\right]&=p_{X}\mu-\frac{1}{2}\sigma_{0}^{2}p_{X}^{2}+\gamma\frac{\partial\mu}{\partial X}-\gamma\mu\frac{\partial\ln{\sigma_{0}}}{\partial X}
\end{align}
\noindent respectively. Note that if $\gamma=0$, then Eq. \eqref{Eq. STS path integral 10} corresponds to the OM functional, while Eq. \eqref{Eq. STS path integral 11} corresponds to the MSRJD functional where $\widetilde{x}=p_{X}$ is called the response field. \vspace{0.25 cm}

\noindent Hence, the OM functional provides a Lagrangian description of the Itô diffusive process, while the MSRJD functional offers a Hamiltonian description of an Itô diffusive process. More generally, Eq. \eqref{Eq. STS path integral 7} and Eq. \eqref{Eq. STS path integral 9} represent the Lagrangian and Hamiltonian descriptions, respectively, regardless of the type of noise in the system. Consequently, the stochastic path integral is extended to any generalized Itô diffusive process. Also, subsequently the action \eqref{Eq. STS path integral 6} will be referred to as the action of generalized Itô diffusive processes (GIDP). \vspace{0.07 cm}

\subsection{\label{Drift-free Stochastic path integral}Additive drift-free stochastic path integral}

\noindent To conclude this section, we consider the special case $\mu\equiv0$ (drift-free), and $\sigma\equiv1$ (additive noise), such that the stochastic path integral for a GIDP (see Eq. \eqref{Eq. STS path integral 6}) becomes (see Appendix \ref{Appendix B})
\begin{equation}
    \label{Eq. STS path integral 12}
    \mathcal{P}_{\eta}^{(0)}(X_{f},t_{f}|X_{0},t_{0})=\int_{-\infty}^{\infty}\frac{dp}{2\pi}e^{ip(X_{f}-X_{0})+\tau\mathcal{K}_{\eta}(p)},
\end{equation}
\noindent where $\tau=t_{f}-t_{0}$. Furthermore, when the CGF depends on $L$ parameters denoted by $\lambda_{1}$, $\lambda_{2}$, $...$, $\lambda_{L}$, it is often the case that this function is scale-invariant with exponents $\nu_{1}$, $\nu_{2}$, $...$, $\nu_{L}$, respectively, i.e.,
\begin{equation}
    \label{Eq. STS path integral 13}
    r\mathcal{K}_{\eta}\left(p;\left\{\lambda_{l}\right\}_{l=1}^{L}\right)= \mathcal{K}_{\eta}\left(p;\left\{\tau^{\nu_{L}}\lambda_{l}\right\}_{l=1}^{L}\right),
\end{equation}
\noindent for all $\tau>0$, and for all $l\in\{1,2,...,L\}$. Thus, if $\mathcal{D}_{\eta}\left(z;\lambda_{1},...,\lambda_{L}\right)\equiv\mathcal{D}_{\eta}\left(z;\left\{\lambda_{l}\right\}_{l=1}^{L}\right)$ is the probability distribution associated with the noise $\eta(t)$ at time $t_{0}$, then the path integral \eqref{Eq. STS path integral 12} becomes
\begin{align}
    \mathcal{P}_{\eta}^{(0)}(X_{f},t_{f}|X_{0},t_{0})&=\int_{\mathbb{R}}\frac{dp}{2\pi}e^{ip(X_{f}-X_{0})+\mathcal{K}_{\eta}\left(p;\left\{\tau^{\nu_{l}}\lambda_{l}\right\}_{l=1}^{L}\right)}\notag\\
    \label{Eq. STS path integral 14}
    &=\mathcal{D}_{\eta}\left(X_{f}-X_{0};\left\{\tau^{\nu_{l}}\lambda_{l}\right\}_{l=1}^{L}\right),
\end{align}
\noindent for all $l\in\{1,2,...,L\}$. In other words, under these conditions, the evolution is completely fixed by the increase in the stochastic variable $X_{f}-X_{0}$, the elapsed time interval $\tau=t_{f}-t_{0}$, the initial fitting parameters $\lambda_{l}$ and the exponents of the scale invariance $\nu_{l}$, for all $l\in\{1,2,...,L\}$. This property will be exploited later to discuss the solutions of the Fokker-Planck equation for a GIDP. Nevertheless, it is unnecessary to mention that Eq. \eqref{Eq. STS path integral 14} is a particular case of a more general relation where $\lambda_{l}$ is transformed into $\Lambda_{l}(\lambda_{1},...,\lambda_{L};t_{f}-t_{0})$ after the time interval $\tau=t_{f}-t_{0}$ has elapsed, for all $l\in\{1,2,...,L\}$. \vspace{0.07 cm}

\section{\label{FP-GIDP}Fokker-Planck equation of generalized Itô diffusive process}

\noindent The path integral for a GIDP stated in equation \eqref{Eq. STS path integral 6}, establishes the transition probability between two configurations of the stochastic variable $X(t)$, but many times due to the functional form of the Hamiltonian (see Equation \eqref{Eq. STS path integral 9}) this is not an easy problem to address. Furthermore, if it is observed that the CGF $\mathcal{K}_{\eta}$ has the same role as the kinetic energy when making an analogy with the Feynman path integral, it follows that the GIDP could be subject to other types of perturbations or restrictions that do not come only from the dynamics of the stochastic drift $\mu\left[X(t),t\right]$ and the diffusion coefficient $\sigma\left[X(t),t\right]$. Thus, it is useful to have an evolution equation for the transition probability $\mathcal{P}_{\eta}\left(X_{f},t_{f}|X_{0},t_{0}\right)$ which will be the objective of this section and which leads to a Fokker-Planck type equation. \vspace{0.25 cm}

\noindent For this purpose, note that the expectation value of any physical quantity $\mathfrak{O}\left[X(t)\right]$ at time $t_{f}$ is given by the transition probability $\mathcal{P}_{\eta}\left(X_{f},t_{f}|X_{0},t_{0}\right)$ since
\begin{equation}
    \label{Eq. Fokker-Planck 1}
    \left\langle\mathfrak{O}\left[ X(t_{f})\right]\right\rangle=\int_{-\infty}^{\infty}\mathcal{P}_{\eta}\left(X_{f},t_{f}|X_{0},t_{0}\right)\mathfrak{O}\left(X_{f}\right)dX_{f},
\end{equation}
\noindent corresponds to a path integral over phase space with a non-fixed endpoint ($X_{f}$ is not fixed). Then, a probability measure $\mathfrak{m}_{0}(dX_{f})$ is established, and defined by
\begin{align}
    \label{Eq. Fokker-Planck 2}
    \mathfrak{m}_{0}(dX_{f}; X_{0},t_{0},t_{f})&=\mathcal{P}_{\eta}\left(X_{f},t_{f}|X_{0},t_{0}\right)dX_{f}.
\end{align}
\noindent Before proceeding, it is important to emphasize that the measure \eqref{Eq. Fokker-Planck 2}, as a probability measure, does not impose any restrictions on the possible values of the stochastic variable $X(t)$ beyond the fixed initial and final conditions of the stochastic path integral. Nonetheless, in practical scenarios, additional constraints often arise. For instance, if $X(t)$ represents the value of a financial derivative, it must remain positive at all times, i.e., $X(t)\geq0$. Similarly, if $X(t)$ corresponds to the position of a particle confined within a box, boundary conditions must be enforced. Hence, to quantify how quickly a GIDP satisfies these constraints, the absorptive rate of a GIDP denoted by $\mathcal{V}\left[X(t),t\right]\equiv\mathcal{V}\left(X,t\right)$ is introduced as
\begin{align}
    \label{Eq. Fokker-Planck 3}
    \mathfrak{m}(dX_{f};X_{0},t_{0},t_{f})&=\int_{X_{0}}^{X_{f}}\int e^{\mathcal{S}_{\eta}+\mathcal{S}_{\eta,I}}\frac{\mathcal{D}p}{2\pi}\mathcal{D}X\;dX_{f},
\end{align}
\noindent where $\mathcal{S}_{\eta,I}=-\int_{t_{0}}^{t_{f}}\mathcal{V} \left[X(\tau),\tau\right]d\tau$. Thus, the function $\mathcal{V}\left[X(t),t\right]\geq0$ quantifies the rate at which the stochastic process $X(t)$ decays when additional constraints are imposed on the system fluctuations. This absorptive rate modifies the probability measure $\mathfrak{m}_{0}(dX_{f}; X_{0},t_{0},t_{f})$ without altering the underlying system dynamics, meaning that the SDE \eqref{Eq. Cumulant Generating Function 2} remains unchanged. Furthermore, it is crucial to emphasize that Eq. \eqref{Eq. Fokker-Planck 3} does not correspond to the introduction of a potential energy term in the Hamiltonian of the system (see Eq. \eqref{Eq. STS path integral 9}), as such a modification would directly impact the evolution of the GIDP (see Eq. \eqref{Eq. Cumulant Generating Function 2}). \vspace{0.25 cm}

\noindent Additionally, it is possible to consider the effect of a source external to the system denoted by $f\left[X(t),t\right]\geq0$ in such a way that the interaction occurs through the absorptive rate with the following expression
\begin{equation}
    \label{Eq. Fokker-Planck 4}
    \small{\mathcal{G}\left(X_{f},t_{f}|X_{0},t_{0}\right) =\int_{t_{0}}^{t_{f}}e^{-\int_{t}^{t_{f}}\mathcal{V}\left[X(\tau),\tau\right]d\tau}f\left[X(t),t\right]dt.}
\end{equation}
\noindent Note that the source $f\left[X(t),t\right]$ in Eq. \eqref{Eq. Fokker-Planck 4} acts as an instantaneous response function, meaning that its interaction with the absorptive rate begins at time $t\in\left[t_{0},t_{f}\right)$ but only takes effect once the system completes its time evolution at $t_{f}$. At this final time, the function $\mathcal{G}\left(X_{f},t_{f}|X_{0},t_{0}\right)$ reaches its maximum value. \vspace{0.25 cm}

\noindent Now, consider the functional
\begin{equation}
    \label{Eq. Fokker-Planck 5}
    \psi\left(X_{f},t_{f}|X_{0},t_{0}\right) =e^{\mathcal{S}_{\eta,I}}+\mathcal{G}\left(X_{f},t_{f}|X_{0},t_{0}\right),
\end{equation}
\noindent hereafter referred to as the Feynman-Kac functional. Thus, if the GIDP evolves from an initial configuration $X(t_{0})=x_{0}$ at a time $t_{0}$, to a final configuration $X(t)=x$ at time $t\in\left[t_{0},t_{f}\right)$, it is satisfied that
\begin{align}
    \label{Eq. Fokker-Planck 6}
    \frac{d}{dt}\psi(x,t|x_{0},t_{0})&=f\left(x,t\right)-\mathcal{V}(x,t)\psi(x,t|x_{0},t_{0}).
\end{align}
\noindent Indeed, Eq. \eqref{Eq. Fokker-Planck 6} shows that $\mathcal{V}\left(x,t\right)$ represents the rate at which the Feynman-Kac functional decays as the source $f\left(x,t\right)$ continues to gain importance in the evolution of the system. \vspace{0.25 cm}

\noindent Consequently, if a differential increase in the expected value of the Feynman-Kac functional is considered (see Appendix \ref{Appendix C}) the following Fokker-Planck type equation is obtained
\begin{widetext}
\begin{equation}
    \label{Eq. Fokker-Planck 7}
    \left[\frac{\partial}{\partial t}+\mu(x,t)\frac{\partial}{\partial x}-\mathcal{K}_{\eta}\left[-i\sigma(x,t)\frac{\partial}{\partial x}\right]+\mathcal{V}(x,t)\right]\Psi(x,t)+\gamma\sigma(x,t)\frac{\partial}{\partial x}\left(\frac{\mu(x,t)}{\sigma(x,t)}\right)\Psi(x,t)=f(x,t),
\end{equation}    
\end{widetext}
\noindent with $\Psi(x,t)=\left\langle\psi(x,t|x_{0},t_{0})\right\rangle$. It is important to highlight Eq. \eqref{Eq. Fokker-Planck 7}, which establishes the general evolution of a GIDP subjected to an absorptive rate $\mathcal{V}\left[X(t),t\right]$ that alters its measure $\mathfrak{m}_{0}$, introduces an external source term to the system denoted by $f\left[X(t),t\right]$, and connects all the different stochastic calculus parameterizations used in the literature through the parameter $\gamma\in\left[0,1\right]$. From now on, this expression is called the Fokker-Planck equation for the GIDP with noise $\eta$ and parameterization $\gamma$ and will be denoted by $FP-GIDP(\eta,\gamma)$. \vspace{0.25 cm}

\noindent The Fokker-Planck equation describes the evolution of a system's probability density through the operator $\mathcal{L}_{FP}$, satisfying $\frac{\partial\Psi}{\partial t}=\mathcal{L}_{FP}\Psi$. Depending on whether $\varepsilon\to0^{+}$ (forward prescription) or $\varepsilon\to0^{-}$ (backward prescription), the equation is governed by $\mathcal{L}_{FP}$ or its adjoint $\mathcal{L}_{FP}^{\dagger}$, respectively \cite{Risken1996}. In this case, the $FP-GIDP(\eta,\gamma)$ corresponds strictly to the forward Fokker-Planck equation for a GIDP with noise $\eta$ and parameterization $\gamma$. \vspace{0.25 cm}

\noindent Hence, the relation $\mathcal{L}_{FP}=-\mathcal{H}_{\eta,I}=-\mathcal{H}_{\eta}-\mathcal{V}(x,t)$ implies a formal solution proportional to $e^{-\int_{t_{0}}^{t}\mathcal{V}\left(x,\tau\right)d\tau}=e^{\mathcal{S}_{\eta,I}}$ (see Eq. \eqref{Eq. Fokker-Planck 3}). This indicates that Eq. \eqref{Eq. Fokker-Planck 7} governs the probability density evolution of a GIDP, with the key distinction being a decay factor proportional to $\mathcal{V}(x,t)$. Additionally, since $\mathcal{H}_{\eta}\sim\mu(x,t)\frac{\partial}{\partial x}$, the backward prescription satisfies $\mathcal{H}_{\eta}^{\dagger}\sim\frac{\partial}{\partial x}\left(\mu(x,t)\Psi(x,t)\right)$ which is a more common term to find in the term associated with stochastic drift in the Fokker-Planck equation. Indeed, note that if $\sigma$ is homogeneous, i.e., $\sigma\equiv\sigma(t)$, and if $\gamma=1$ (Hänggi-Klimontovich integral), then the second and fifth terms on the left side of the $FP-GIDP(\eta,\gamma)$ reduce to
\begin{align*}
    \mu\frac{\partial\Psi}{\partial x}+\gamma\Psi\sigma\frac{\partial}{\partial x}\left(\sigma^{-1}\mu\right)=\mu\frac{\partial\Psi}{\partial x}+\Psi\frac{\partial\mu}{\partial x}=\frac{\partial}{\partial x}\left(\mu\Psi\right).
\end{align*}

\noindent Furthermore, it is important to highlight the role of $\Psi(x,t)$ as an expectation value of the Feynman-Kac functional (see Eq. \eqref{Eq. Fokker-Planck 5}), since in the particular case where $\eta(t)$ represents a white noise with stochastic drift $\mu\equiv\mu_{0}(x,t)$, and diffusion coefficient $\sigma\equiv\sigma_{0}\left(x,t\right)$, the $FP-GIDP(\eta,\gamma)$ reproduces the well-known result of the Feynman-Kac formula with $\gamma=0$, namely
\begin{equation}
    \label{Eq. Fokker-Planck 8}
    \left[\frac{\partial}{\partial t}+\mu\frac{\partial}{\partial x}-\frac{\sigma^{2}}{2}\frac{\partial^{2}}{\partial x^{2}}+\mathcal{V}(x,t)\right]\Psi(x,t)=f(x,t).
\end{equation}
\noindent The main difference of the Feynman-Kac formula of Eq. \eqref{Eq. Fokker-Planck 8} with that found in the literature lies in the selection of the sign of $\mathcal{K}_{\eta}(\sigma p)$, $\mathcal{V}(x,t)$ and $f(x,t)$. This difference lies in the integration limits chosen for Feynman-Kac functional (see Eq. \eqref{Eq. Fokker-Planck 5}), which is linked to the fact that the Feynman-Kac formula is thought of as a backward Fokker-Planck equation with a final condition instead of an initial condition \cite{Risken1996}. Thus, without loss of generality, the initial condition $\Psi(x,t_{0})=\delta(x-x_{0})$ is taken, which indicates that at the beginning of the evolution of the GIDP the value of the stochastic variable $X(t)$ is known exactly. \vspace{0.25 cm}

\noindent Now, note that in specific cases where the stochastic drift $\mu(x,t)$ and the diffusion coefficient $\sigma(x,t)$ remain constant, the $FP-GIDP(\eta,\gamma)$ becomes independent of the parameterization $\gamma\in\left[0,1\right]$. More generally, if the drift term $\mu(x,t)$ is proportional to the diffusion coefficient $\sigma(x,t)$, such that $\mu(x,t)=C\sigma(x,t)$ with $C\in\mathbb{R}$, the $FP-GIDP(\eta,\gamma)$, also loses dependence on $\gamma\in\left[0,1\right]$ since $\frac{\partial}{\partial x}\left(\sigma^{-1}\mu\right)=0$ (see the fifth term on the left side of Eq. \eqref{Eq. Fokker-Planck 7}). This condition is particularly relevant in geometric Brownian motion, where $\mu(x,t) = \mu_{0}x$ and $\sigma(x,t)=\sigma_{0}x$, with $\mu_{0}\in\mathbb{R}$ and $\sigma_{0}>0$. The proportionality between drift and diffusion terms ensures scale invariance and parameterization independence, making geometric Brownian motion a fundamental model in financial mathematics and stochastic modeling \cite{Giordano2023, Stojkoski2020}. \vspace{0.25 cm}

\noindent Nevertheless, when multiple noise sources influence a stochastic process $X(t)$, the conditions for parameterization independence can vary significantly. For instance, introducing two correlated Gaussian white noises can induce non-equilibrium transitions in stochastic systems \cite{Denisov2003} or lead to different stability regimes in harmonic oscillators subjected to additive and multiplicative noise \cite{Mendez2014}. Additionally, some generalizations of geometric Brownian motion incorporating nonlinear drift or heterogeneous diffusion can result in more complex conditions where independence from the stochastic calculus parameterization is harder to achieve \cite{Giordano2023, Dupont2024}. \vspace{0.07 cm}

\subsection{\label{Full Stochastic path integral}Fokker-Planck equation for homogeneous and constant coefficients}

\noindent Finally, to consider a practical and frequent example in the literature, consider that the stochastic drift and the diffusion coefficient are constant and homogeneous functions, that is, $\mu(x,t)=\mu_{0}$ and $\sigma(x,t)=\sigma_{0}>0$. Also, the source term is null ($f(x,t)\equiv0$), and the absorptive rate is constant and equal to $V_{0}\geq0$. In this case, taking the Fourier transform of Eq. \eqref{Eq. Fokker-Planck 7} with respect to the variable $x$, it is verified that its general solution of the $FP-GIDP(\eta,\gamma)$ is 
\begin{equation}
    \label{Eq. Fokker-Planck 9}
    \Psi(x,t)=\frac{e^{-\tau V_{0}}}{\sigma_{0}}\int_{-\infty}^{\infty}e^{ipz_{x}+\tau\mathcal{K}_{\eta}\left(p\right)}\frac{dp}{2\pi},
\end{equation}
\noindent where $\tau=t-t_{0}$, and $z_{x}=(x-x_{0}-\mu_{0}\tau)/\sigma_{0}$, such that for $t=t_{0}$ we have $\Psi(x,t_{0})=\delta(x-x_{0})$. Hence, in the particular case where the CGF depends on $L$ parameters denoted by $\lambda_{1}$, $\lambda_{2}$, $...$, $\lambda_{L}$ that satisfy Eq. \eqref{Eq. STS path integral 13}, then the solution of Eq. \eqref{Eq. Fokker-Planck 8} satisfies by Eq. \eqref{Eq. STS path integral 14} that
\begin{equation}
    \label{Eq. Fokker-Planck 10}
    \Psi(x,t)=\frac{e^{-\tau V_{0}}}{\sigma_{0}}\mathcal{D}_{\eta}\left(z_{x};\tau^{\nu_{1}}\lambda_{1},...,\tau^{\nu_{L}}\lambda_{L}\right).
\end{equation}
\noindent More generally, consider the case where the $FP-GIDP(\eta,\gamma)$ is independent of $\gamma\in[0,1]$, i.e. $\sigma(x,t)=\rho_{0}\mu(x,t)$, with $\rho_{0}\in\mathbb{R}$ and $\mu(x,t)$ a strictly increasing continuously differentiable function respect a $x$, and the same conditions as above apply on $V(x,t)$ and $f(x,t)$. Thence, with change of variable $y(x)=\int_{x_{0}}^{x}\mu^{-1}(\xi,t)d\xi$, we have that $\frac{\partial}{\partial y}=\mu(x,t)\frac{\partial}{\partial x}$, and Eq. \eqref{Eq. Fokker-Planck 7} becomes an equation with constant coefficients and in such case, the solution of the equation is
\begin{equation}
    \label{Eq. Fokker-Planck 11}
    \Psi(y,t)=\frac{e^{-\tau V_{0}}}{\sigma(x,t)}\int_{-\infty}^{\infty}e^{ipw_{x}+\tau\mathcal{K}_{\eta}\left(p\right)}\frac{dp}{2\pi},
\end{equation}
\noindent where $w_{x}=(y(x)-y(x_{0})-\tau)/\rho_{0}$, and the factor $\sigma^{-1}(x,t)=\rho_{0}^{-1}\mu^{-1}(x,t)=\rho_{0}^{-1}\frac{dy}{dx}$ outside the integral ensures that for $t=t_{0}$ we have $\Psi(x,t_{0})=\delta(x-x_{0})$. \vspace{0.25 cm}

\noindent Note that in both cases (Eq. \eqref{Eq. Fokker-Planck 9} and Eq. \eqref{Eq. Fokker-Planck 11}), the average value of the Feynman-Kac functional corresponds to the probability density of the process $X(t)$, with the essential difference that the distribution decays more rapidly in time due to the effect of the absorption rate $V(x,t)=V_{0}\geq0$. \vspace{0.25 cm}

\noindent As a final note for this section, it is important to emphasize that the term associated with stochastic drift can also take the form of a memory kernel, such as $\int_{t_{0}}^{t}M(\tau)X(\tau)d\tau$. In particular, when $M(\tau)=\frac{1}{\Gamma(1-\alpha)}(t-\tau)^{-\alpha}$ with $0<\alpha<1$, the integral is interpreted as a fractional operator \cite{Quevedo2024}. This implies that the GIDP acquires a long-range memory, even though such memory was not assumed during the derivation of Eq. \eqref{Eq. Fokker-Planck 7}. This is because it was only assumed that the stochastic drift $\mu\left[X(t),t\right]$, the diffusive coefficient $\sigma\left[X(t),t\right]$, the absorptive rate $\mathcal{V}\left[X(t),t\right]$, and the source term $f\left[X(t),t\right]$ are continuous functions almost everywhere within the interval $\left[t_{0},t\right]$, ensuring their Riemann-integrability in time. \vspace{0.25 cm}

\noindent This result is significant as it demonstrates that the additivity of evolution over different times in the evolution operator $\mathcal{H}_{\eta,I}$ of the stochastic path integral depends primarily on the continuity of the functions $\mu(x,t)$, $\sigma(x,t)$, $\mathcal{V}(x,t)$, and $f(x,t)$, as well as on the continuity of the noise trajectories $\eta(t)$, even when they exhibit abrupt jumps, such as in a Lévy flight process. This is more relevant than the Markovian property of some stochastic processes. Nevertheless, when the stochastic drift is represented by a fractional or long-range operator, such as a convolution, the discretization in Appendix \ref{Appendix A} should be handled carefully to ensure the correct estimation of the system Jacobian. Furthermore, a fundamentally different case (where significant modifications to this stochastic path integral formulation would be expected) is when the GIDP is driven by an underlying noise governed by a fractional operator, i.e., considering $\frac{d^{\alpha}X}{dt^{\alpha}}$, with $\alpha>0$, and $\alpha\neq1$ instead of $\frac{dX}{dt}$, since this would also alter the system Jacobian and the action of the stochastic path integral. \vspace{0.07 cm}

\section{\label{Stochastic Processes application}Typical stochastic processes with threshold}

\noindent From the formulation developed so far for GIDPs, several of the classical stochastic processes from the literature can be developed with this approach and to show the potentiality of the $FP-GIDP(\eta,\gamma)$, we consider $4$ types of stochastic processes: Brownian motion ($BM$), geometric Brownian motion ($GBM$), Levy $\alpha$-stable flight ($LF(\alpha)$), and geometric Levy $\alpha$-stable flight ($GLF(\alpha)$). Also, we establish a boundary condition on these $4$ stochastic processes such that if $X(t)<x_{V}$, then $X(t)=0$, where $x_{V}$ is a threshold imposing the boundary condition $\Psi(x,t)=0$ if $x<x_{V}$. It is important to note that the threshold $x_{V}$ plays a crucial role in modeling real-world constraints on the stochastic variable $X(t)$. For example, in population dynamics or financial markets, where variables must remain non-negative, a threshold at zero ensures consistency with these natural constraints. Similarly, in physical systems, such as the motion of a particle confined within a box, thresholds define the permissible range of positions. In general, $x_{V}$ serves as a lower or upper bound that keeps the modeled process within realistic limits. \vspace{0.25 cm}

\noindent To incorporate this threshold as a boundary condition in the stochastic path integral for a GIDP, we use the absorption rate $\mathcal{V}(x,t)$ to describe how quickly the stochastic process $X(t)$ decays over time (see Eq. \eqref{Eq. Fokker-Planck 11}). Specifically, we set $\mathcal{V}(x,t)=V_{0}\mathbb{H}(x-x_{V})$, where $V_{0}\to\infty$ and $\mathbb{H}$ is the Heaviside function, ensuring that $\Psi(x,t)=0$ for $x<x_{V}$, effectively removing those trajectories. This behaves similarly to an infinite potential barrier in quantum mechanics, preventing $X(t)$ from crossing into the forbidden region. However, unlike a reflective boundary, which forces $X(t)$ to reverse upon reaching $x_{V}$, the absorbing boundary eliminates trajectories reaching this threshold. In the case of the four stochastic processes discussed earlier, $x_{V}$ is an arbitrary real value for $BM$ and $LF(\alpha)$, while for $GBM$ and $GLF(\alpha)$, it is a non-negative value due to the inherent constraints imposed by the governing equations where $X(t)$ only admits non-negative values for a fixed $X_{0}\geq0$. \vspace{0.25 cm}

\noindent Now, the CGF of $BM\equiv BM(\nu,\rho)$, and $LF(\alpha)\equiv LF(\alpha,\beta,\nu,\rho)$ are defined by the following expressions
\begin{align}
    \label{Eq. Entropy Production 1}
    \mathcal{K}_{BM}(q;\nu,\rho)&=iq\nu+\frac{\rho^{2}}{2}(iq)^{2},\\
    \label{Eq. Entropy Production 2}
    \mathcal{K}_{LF}(q;\alpha,\beta,\nu,\rho)&=iq\nu-\left|\rho q\right|^{\alpha}\left(1-i\beta\frac{q}{|q|}\Phi(\alpha)\right),
\end{align}
\noindent respectively, where $\nu\in\mathbb{R}$ is the localization parameter, $\rho>0$ is the scale parameter, $\beta\in[-1,1]$ is the skewness parameter, and $\alpha\in(0,2]$ is the stability parameter, and $\Phi(\alpha)=tan{\left(\frac{\pi\alpha}{2}\right)}$ for $\alpha\neq1$. From the expressions \eqref{Eq. Entropy Production 1} and \eqref{Eq. Entropy Production 2}, it is clear that
\begin{align}
    \label{Eq. Entropy Production 3}
    \tau\mathcal{K}_{BM}(q;\nu,\rho)&=\mathcal{K}_{BM}(q;\nu\tau,\rho\sqrt{\tau}),\\
    \label{Eq. Entropy Production 4}
    \tau\mathcal{K}_{LF}(q;\alpha,\beta,\nu,\rho)&=\mathcal{K}_{LF}(q;\alpha,\beta,\nu\tau,\rho\tau^{1/\alpha}),
\end{align}
\noindent for all $\tau>0$. Thus, to denote that a noise $\eta(t)$ satisfies these CGF it will be denoted by $\eta_{BM}(t;\nu,\rho)$ or $\eta_{LF}(t;\alpha,\beta,\nu,\rho)$. \vspace{0.25 cm}

\noindent On the other hand, $BM$, $LF(\alpha)$, $GBM$, and $GLF(\alpha)$ are formally defined as GIDP through the following stochastic differential equations
\begin{align}
    \label{Eq. Entropy Production 5}
    \dot{X}_{BM}(t)&=\mu_{0}+\sigma_{0}\eta_{BM}(t;0,1),\\
    \label{Eq. Entropy Production 6}
    \dot{X}_{LF}(t)&=\mu_{0}+\sigma_{0}\eta_{LF}(t;\alpha,\beta,0,1),\\
    \label{Eq. Entropy Production 7}
    \dot{X}_{GBM}(t)&=\mu_{0}X(t)+\sigma_{0}X(t)\eta_{BM}(t;0,\rho),\\
    \label{Eq. Entropy Production 8}
    \dot{X}_{GLF}(t)&=\mu_{0}X(t)+\sigma_{0}X(t)\eta_{LF}(t;\alpha,\beta,0,1),
\end{align}
\noindent respectively. Therefore, in any of these $4$ cases, we are in the conditions of the expressions \eqref{Eq. Fokker-Planck 10} and \eqref{Eq. Fokker-Planck 11}, which imply general solutions of the form $\Psi(x,t)=\mathcal{N}\sigma_{0}^{-1}\mathcal{D}_{\eta}\left(z_{x};\tau^{\nu_{1}}\lambda_{1},...,\tau^{\nu_{L}}\lambda_{L}\right)$, where $\mathcal{N}$ is a normalization constant that satisfies
\begin{equation}
    \label{Eq. Entropy Production 9}
    \mathcal{N}^{-1}=1-\int_{-\infty}^{z_{V}}\mathcal{D}_{\eta}\left(x;\tau^{\nu_{1}}\lambda_{1},...,\tau^{\nu_{L}}\lambda_{L}\right)dx,
\end{equation}
\noindent with $z_{V}=\sigma_{0}^{-1}(x_{V}-x_{0}-\mu_{0}\tau)$, since the normalization condition Eq. \eqref{Eq. Cumulant Generating Function 6} is satisfied for $\sigma_{0}^{-1}\mathcal{D}_{\eta}\left(z_{x};\cdot\right)$. Note that the case $\sigma(x,t)=\rho_{0}\mu(x,t)$ is also contemplated in the general solution $\Psi(x,t)=\mathcal{N}\sigma_{0}^{-1}\mathcal{D}_{\eta}\left(z_{x};\tau^{\nu_{1}}\lambda_{1},...,\tau^{\nu_{L}}\lambda_{L}\right)$ since $\sigma^{-1}(x,t)=\rho_{0}^{-1}\frac{dy}{dx}$ allows us to return to the same situation but with the integral over the variable $y=y(x)$. \vspace{0.25 cm}

\noindent Hence, the normalization constant corresponds to the complement of the cumulative distribution function of the variable $\eta(t)$ evaluated in $z_{V}$, that is, the survival function of $\eta(t)$ evaluated in $z_{V}$. Consequently, the solutions of the stochastic processes \eqref{Eq. Entropy Production 5}, \eqref{Eq. Entropy Production 6}, \eqref{Eq. Entropy Production 7}, and \eqref{Eq. Entropy Production 8} are
\begin{align}
    \label{Eq. Entropy Production 10}
    \Psi_{BM}(x,t)&=\frac{\mathfrak{f}_{BM}\left(z(x);0,\sqrt{\tau}\right)}{\sigma_{0}\;\mathfrak{F}_{BM}\left(z_{V};0,\sqrt{\tau}\right)},\\
    \label{Eq. Entropy Production 11}
    \Psi_{LF}(x,t)&=\frac{\mathfrak{f}_{LF}\left(z(x);\alpha,\beta,0,\tau^{1/\alpha}\right)}{\sigma_{0}\;\mathfrak{F}_{LF}\left(z_{V};\alpha,\beta,0,\tau^{1/\alpha}\right)},\\
    \label{Eq. Entropy Production 12}
    \Psi_{GBM}(x,t)&=\frac{\mathfrak{f}_{BM}\left(w(x);0,\sqrt{\tau}\right)}{\sigma_{0}x\;\mathfrak{F}_{BM}\left(w_{V};0,\sqrt{\tau}\right)},\\
    \label{Eq. Entropy Production 13}
    \Psi_{GLF}(x,t)&=\frac{\mathfrak{f}_{LF}\left(w(x);\alpha,\beta,0,\tau^{1/\alpha}\right)}{\sigma_{0}x\;\mathfrak{F}_{LF}\left(w_{V};\alpha,\beta,0,\tau^{1/\alpha}\right)},
\end{align}
\noindent respectively, where $\sigma_{0}z(x)=x-x_{0}-\mu_{0}(t-t_{0})$, $\sigma_{0}w(x)=\ln{x}-\ln{x_{0}}-\mu_{0}(t-t_{0})$, $z_{V}=z(x=x_{V})$, $w_{V}=w(x=x_{V})$, $\mathfrak{f}_{BM}(\xi;0,\sqrt{\tau}))$ and $\mathfrak{f}_{LF}(\xi;\alpha,\beta,0,\tau^{1/\alpha}))$ are the probability density function of Brownian motion and Levy $\alpha$-stable flight, and $\mathfrak{F}_{BM}\left(z;0,\sqrt{\tau}\right)$ and $\mathfrak{F}_{LF}\left(z;\alpha,\beta,0,\tau^{1/\alpha}\right)$ are their respective survival functions defined by
\begin{align}
    \label{Eq. Entropy Production 14}
    \mathfrak{F}_{BM}(z)&=\int_{z}^{\infty}\mathfrak{f}_{BM}(\xi;0,\sqrt{\tau})d\xi=\int_{z}^{\infty}\frac{e^{-\frac{\xi^{2}}{2\tau}}}{\sqrt{2\pi\tau}}d\xi,\\
    \label{Eq. Entropy Production 15}
    \mathfrak{F}_{LF}(z)&=\int_{z}^{\infty}\mathfrak{f}_{LF}\left(\xi;\alpha,\beta,0,\tau^{1/\alpha}\right)d\xi.
\end{align}
\noindent In the following subsections, the validity and scope of the expressions \eqref{Eq. Entropy Production 10}, \eqref{Eq. Entropy Production 11}, \eqref{Eq. Entropy Production 12}, and \eqref{Eq. Entropy Production 13} are explored for the four stochastic processes under consideration. Note that the usual case where $x_{V}\to-\infty$ implies the usual dynamics in the $BM$ and $LF(\alpha)$ since $z_{V}\to-\infty$ and $\mathcal{N}=1$ (see Eq. \eqref{Eq. Entropy Production 9}), while $x_{V}\to0^{+}$ implies the simplest dynamics in the $GBM$ and $GLF(\alpha)$ since $w_{V}\sim\ln{x_{V}}\to-\infty$ and $\mathcal{N}=1$. Therefore, in the last two cases a condition $x_{V}\geq1$ is taken to explore the effects of the threshold on the probability density function of these two stochastic processes. \vspace{0.25 cm}

\noindent Finally, a key aspect is the behavior of noise as a function of infinitesimal time increments, which is crucial for numerical stochastic integration. The significance of expressions \eqref{Eq. Entropy Production 3} and \eqref{Eq. Entropy Production 4} lies in their guidance on performing stochastic integration when the time increment $\tau\to0^{+}$ is applied to processes such as Brownian motion and Levy $\alpha$-stable flights. In the simulations, these expressions ensure that the noise behaves appropriately, reflecting the system's dynamics over small time scales. \vspace{0.25 cm}

\noindent Additionally, please note that the simulation of stochastic processes in \eqref{Eq. Entropy Production 5}, \eqref{Eq. Entropy Production 6}, \eqref{Eq. Entropy Production 7}, and \eqref{Eq. Entropy Production 8} was carried out using the Euler-Maruyama algorithm \cite{Kloeden1992}. To ensure the accuracy of these simulations, expressions \eqref{Eq. Entropy Production 3} and \eqref{Eq. Entropy Production 4} were applied when considering temporal increments of the noise $\eta(t)$. It is worth mentioning that as time steps decrease to better capture stochastic behavior, the number of iterations in the Euler-Maruyama method increases, which can cause a bottleneck in computational performance. Therefore, the codes implemented are parallelized to make each simulation as efficient as possible. Readers interested in reproducing or extending these simulations can access the associated code through a GitHub repository, which provides the necessary scripts and tools for further exploration \cite{GithubFPD2023}. \vspace{0.07 cm}

\subsection{\label{Evolution PDF}Evolution of the probability density function of the stochastic processes with threshold}

\noindent To compare the expressions \eqref{Eq. Entropy Production 10}, \eqref{Eq. Entropy Production 11}, \eqref{Eq. Entropy Production 12}, and \eqref{Eq. Entropy Production 13} with numerical simulations, it is important to mention that from now on $N_{s}$ denotes the number of simulations performed in each of the four stochastic processes addressed, $N_{t}$ the number of time steps considered in each simulation, $N_{b}$ the number of bins used to group the data of $\Psi(x,t)$, and $t_{f}$ the final time instant until which the system is made to evolve. Furthermore, note that $\Psi(x,t)$ is interpreted as the probability density function for each of the four stochastic processes studied since it satisfies the normalization condition Eq. \eqref{Eq. Cumulant Generating Function 6} and is written in terms of the noise probability density function $\mathcal{D}_{\eta}(\cdot)$ with the substantial difference being the rapid decay caused by the factor $e^{-\tau V_{0}}$ when $V_{0}>0$ is finite (see Eq. \eqref{Eq. Fokker-Planck 10} and Eq. \eqref{Eq. Fokker-Planck 11}). \vspace{0.25 cm}

\begin{figure*}
    \centering
    \includegraphics[scale=0.295]{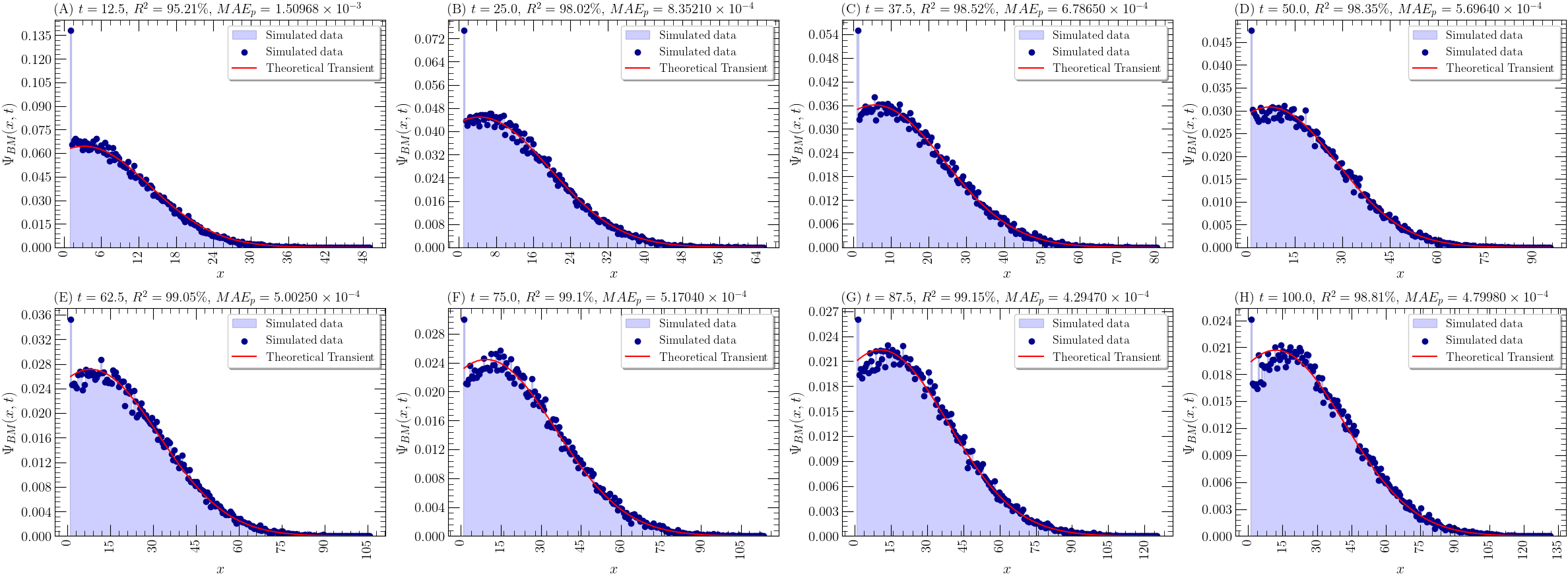}
    \caption{Temporal Evolution of the probability density function $\Psi_{BM}(x,t)$ for restricted Brownian Motion with parameters $x_{0}=2$, $t_{0}=0$, $\mu=1\times10^{-1}$, $\sigma=3$, and $x_{V}=1$, using $N_{s}=4\times10^{4}$ trajectories, $N_{t}=5\times10^{3}$ time steps, $N_{b}=2\times10^{2}$ bins, and a final time of $t_{f}=1\times10^{2}$. In all cases, the solid line corresponds to the theoretical fit while the points correspond to the simulated data.}
    \label{Fig_1.0}
\end{figure*}

\noindent Figure \ref{Fig_1.0} illustrates the temporal evolution of the probability density function for restricted Brownian motion (see Eq. \eqref{Eq. Entropy Production 5}) based on $N_{s}=4\times10^{4}$ trajectories, $N_{t}=5\times10^{3}$ time steps, $N_{b}=2\times10^{2}$ bins, and a final time of $t_{f}=1\times10^{2}$. The remaining parameters—initial value, initial time, stochastic drift, diffusion coefficient, and threshold value—are set to $x_{0}=2$, $t_{0}=0$, $\mu=1\times10^{-1}$, $\sigma=3$, and $x_{V}=1$, respectively. Comparing the results at eight different time instants, Eq. \eqref{Eq. Entropy Production 10} (solid red line) shows a precise fit with the simulated data (points and histogram), validating the analytical solution. Indeed, Table \ref{Table_1.1} further presents the coefficients of determination $R^{2}$ and the mean absolute error $MAE_{1}$ for each time instant. All $R^{2}$ values exceed $95.21\%$, confirming the accuracy of the fit, while the $MAE_{1}$ values demonstrate low variability between the simulated data and the theoretical fit, given that the distribution's peak is around $x_{0}-\mu\tau$. Additionally, a condensation effect is observed near the threshold value $x_{V}$ early in the evolution, which dissipates over time. This indicates a broader range of states for the stochastic variable $X(t)$ while still satisfying the condition $\Psi(x,t)=0$ for all $x<x_{V}$. \vspace{0.25 cm}

\begin{figure*}
    \centering
    \includegraphics[scale=0.295]{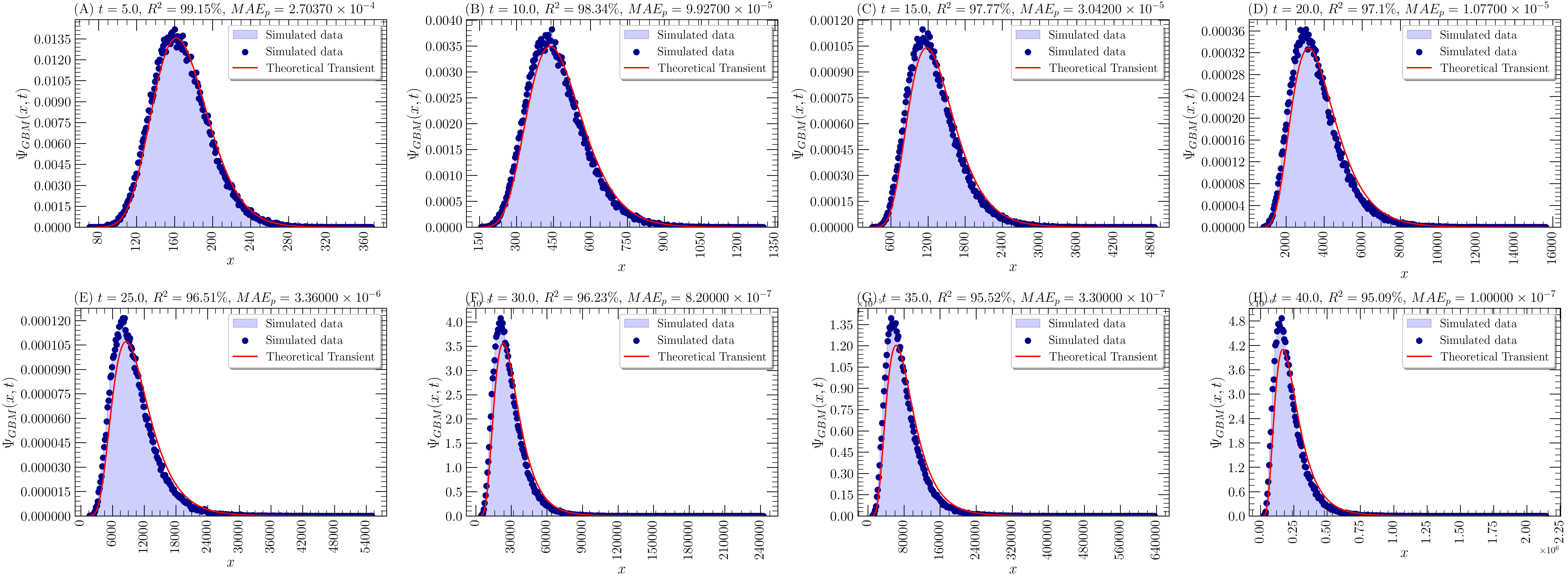}
    \caption{Temporal Evolution of the probability density function $\Psi_{GBM}(x,t)$ for restricted geometric Brownian Motion with parameters $x_{0}=6\times10^{1}$, $t_{0}=0$, $\mu=2.05\times10^{-1}$, $\sigma=8\times10^{-2}$, and $x_{V}=1$, using $N_{s}=1\times10^{5}$ trajectories, $N_{t}=4\times10^{3}$ time steps, $N_{b}=4\times10^{2}$ bins, and a final time of $t_{f}=4\times10^{1}$. In all cases, the solid line corresponds to the theoretical fit while the points correspond to the simulated data.}
    \label{Fig_1.1}
\end{figure*}

\noindent Figure \ref{Fig_1.1} illustrates the temporal evolution of the probability density function for restricted geometric Brownian motion (see Eq. \eqref{Eq. Entropy Production 6}) based on $N_{s}=1\times10^{5}$ trajectories, $N_{t}=4\times10^{3}$ time steps, $N_{b}=4\times10^{2}$ bins, and a final time of $t_{f}=4\times10^{1}$. The remaining parameters—initial value, initial time, stochastic drift, diffusion coefficient, and threshold value—are set to $x_{0}=6\times10^{1}$, $t_{0}=0$, $\mu=2.05\times10^{-1}$, $\sigma=8\times10^{-2}$, and $x_{V}=1$, respectively. By comparing the simulation results at eight different time points, the accuracy of equation \eqref{Eq. Entropy Production 11} (solid red line) is confirmed against the simulated data (points and histogram).  Furthermore, a rapid increase in the peak of the distribution is observed, along with a broader range of possible values for the stochastic variable $X(t)$, indicating faster average growth compared to the restricted $BM$. This is consistent with the fact that for standard $GBM$, $\mathbb{E}\left[X(t)\right]=x_{0}e^{\mu_{0}\tau}$, implying a higher growth rate than the typical $x_{0}+\mu\tau$ for standard $BM$. Thence, for longer times, the numerical fit requires more simulations for higher precision around the peak, as illustrated in panel $(H)$ of Figure \ref{Fig_1.1}, where $t=t_{f}$. Nevertheless, the fit is quite accurate, as shown in Table \ref{Table_1.1} with the coefficients of determination $R^{2}$ and the mean absolute error $MAE_{1}$ for each time instant. All values of $R^{2}$ exceed $95.09\%$, while the $MAE_{1}$ values exhibit even lower variability than those for the restricted $BM$. \vspace{0.25 cm}

\begin{table*}
\centering
\caption{Coefficients of determination $R^{2}$ and mean absolute error $MAE_{1}$ for the simulations made of restricted $BM$ and restricted $GBM$ at different time instants. The status column corresponds to the literal indicated in Figure \ref{Fig_1.0} and Figure \ref{Fig_1.1}.}
\label{Table_1.1}
\begin{tabular}{@{}ccccccc@{}}
\toprule
Status & $t_{BM}$ & $R_{BM}^{2}(\%)$ & $MAE_{1}^{(BM)}(\times10^{-4})$ & $t_{GBM}$ & $R_{GBM}^{2}(\%)$ & $MAE_{1}^{(GBM)}(\times10^{-5})$ \\ \midrule
A & $12.5$ & $95.21$ & $15.1$ & $5.0$ & $99.15$ & $27.00$\\
B & $25.0$ & $98.02$ & $8.35$ & $10.0$ & $98.34$ & $9.930$\\
C & $37.5$ & $98.52$ & $6.79$ & $15.0$ & $97.77$ & $3.040$\\
D & $50.0$ & $98.35$ & $5.70$ & $20.0$ & $97.10$ & $1.080$\\
E & $62.5$ & $99.05$ & $5.00$ & $25.0$ & $96.51$ & $0.336$\\
F & $75.0$ & $99.10$ & $5.17$ & $30.0$ & $96.23$ & $0.082$\\
G & $87.5$ & $99.15$ & $4.29$ & $35.0$ & $95.52$ & $0.033$\\
H & $100.0$ & $98.81$ & $4.80$ & $40.0$ & $95.09$ & $0.011$\\
\bottomrule
\end{tabular}
\end{table*}

\begin{figure*}
    \centering
    \includegraphics[scale=0.295]{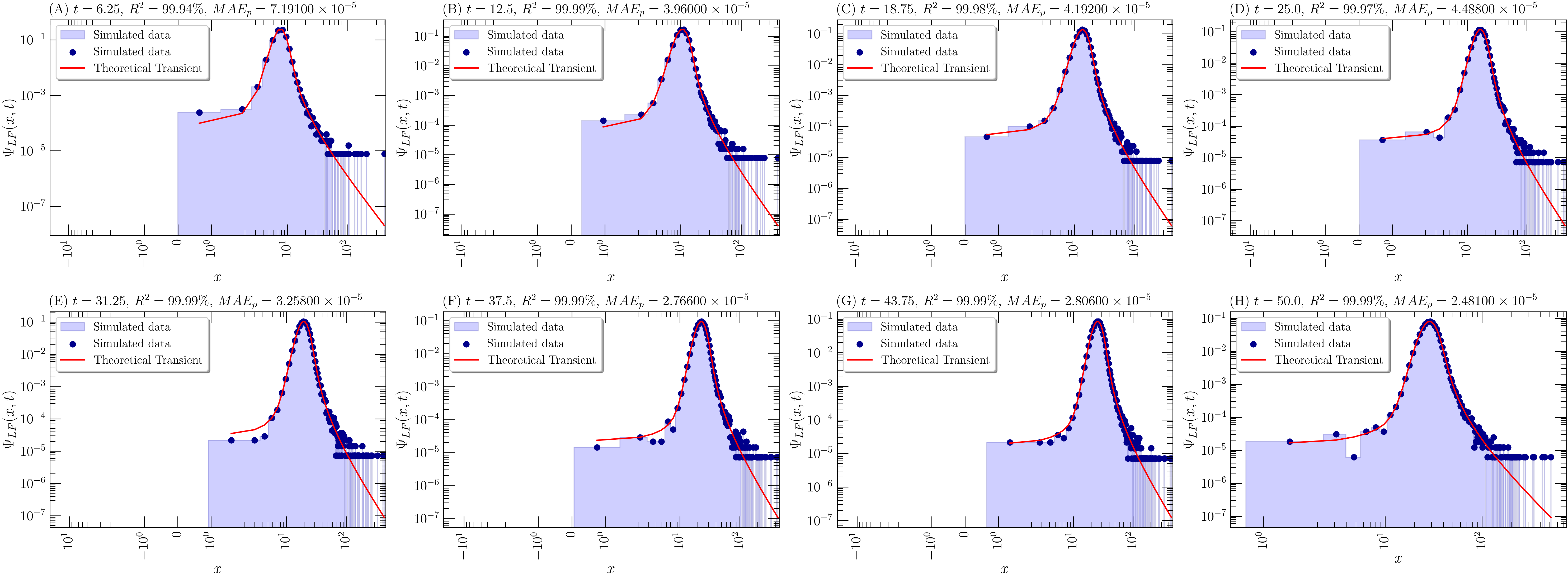}
    \caption{Temporal Evolution of the probability density function $\Psi_{LF}(x,t)$ for restricted Levy $\alpha$-stable flight with parameters $x_{0}=5$, $t_{0}=0$, $\alpha=1.8$, $\beta=0.9$, $\mu=5\times10^{-1}$, $\sigma=4\times10^{-1}$, and $x_{V}=-1$, using $N_{s}=1\times10^{5}$ trajectories, $N_{t}=5\times10^{3}$ time steps, $N_{b}=2\times10^{2}$ bins, and a final time of $t_{f}=5\times10^{1}$. In all cases, the solid line corresponds to the theoretical fit while the points correspond to the simulated data.}
    \label{Fig_1.2}
\end{figure*}

\begin{figure*}
    \centering
    \includegraphics[scale=0.295]{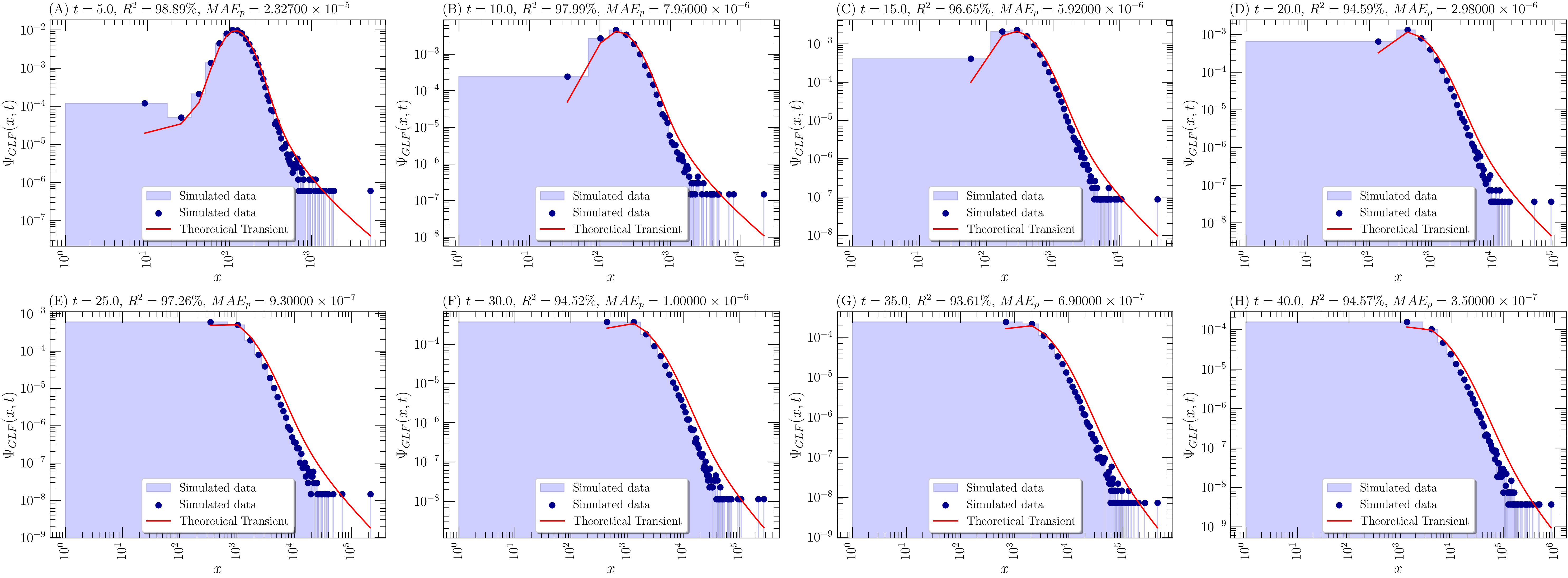}
    \caption{Temporal Evolution of the probability density function $\Psi_{GLF}(x,t)$ for restricted geometric Levy $\alpha$-stable flight with parameters $x_{0}=8\times10^{1}$, $t_{0}=0$, $\alpha=1.9$, $\beta=0.5$, $\mu=1.05\times10^{-1}$, $\sigma=1\times10^{-1}$, and $x_{V}=1$, using $N_{s}=1\times10^{5}$ trajectories, $N_{t}=8\times10^{3}$ time steps, $N_{b}=2\times10^{2}$ bins, and a final time of $t_{f}=4\times10^{1}$. In all cases, the solid line corresponds to the theoretical fit while the points correspond to the simulated data.}
    \label{Fig_1.3}
\end{figure*}

\noindent Now, repeating the process with Levy $\alpha$-stable flights, Figure \ref{Fig_1.2} illustrates the temporal evolution of the probability density function for restricted Levy $\alpha$-stable flight (see Eq. \eqref{Eq. Entropy Production 7}) based on $N_{s}=1\times10^{5}$ trajectories, $N_{t}=5\times10^{3}$ time steps, $N_{b}=2\times10^{2}$ bins, and a final time of $t_{f}=5\times10^{1}$. The remaining parameters—initial value, initial time, stability parameter, skewness parameter, stochastic drift, diffusion coefficient, and threshold value—are set to $x_{0}=5$, $t_{0}=0$, $\alpha=1.8$, $\beta=0.9$, $\mu=5\times10^{-1}$, $\sigma=4\times10^{-1}$, and $x_{V}=-1$, respectively. Also, Figure \ref{Fig_1.3} shows the temporal evolution of the probability density function for restricted geometric Levy $\alpha$-stable flight (see Eq. \eqref{Eq. Entropy Production 8}) based on $N_{s}=1\times10^{5}$ trajectories, $N_{t}=8\times10^{3}$ time steps, $N_{b}=2\times10^{2}$ bins, and a final time of $t_{f}=4\times10^{1}$. The remaining parameters—initial value, initial time, stability parameter, skewness parameter, stochastic drift, diffusion coefficient, and threshold value—are set to $x_{0}=8\times10^{1}$, $t_{0}=0$, $\alpha=1.9$, $\beta=0.5$, $\mu=1.05\times10^{-1}$, $\sigma=1\times10^{-1}$, and $x_{V}=1$, respectively. In both cases, comparing the results of the simulations at eight different time points confirms the accuracy of Eq. \eqref{Eq. Entropy Production 12} and Eq. \eqref{Eq. Entropy Production 13} (solid red lines) against the simulated data (dots and histograms). \vspace{0.25 cm}

\noindent Additionally, it is important to note that Figures \ref{Fig_1.2} and \ref{Fig_1.3} utilize a logarithmic scale to highlight the power laws characteristic of a Lévy $\alpha$-stable flight. Due to the large jumps inherent in Levy $\alpha$-stable flight distributions, Figure \ref{Fig_1.2} shows that the condensation effect around the threshold $x_{V}$ is less pronounced than the restricted $BM$ case. Similarly, the sharp peak observed in restricted $GBM$ is less prominent in the restricted $GLF(\alpha)$, as the large jumps allow the stochastic variable to spread more uniformly across the logarithmic scale, especially evident in panels $(D)-(H)$ of Figure \ref{Fig_1.3}. Finally, it is worth mentioning that the coefficients of determination $R^{2}$ are greater than $99.94\%$ for the restricted $LF(\alpha)$ and greater than $93.61\%$ for the restricted $GLF(\alpha)$, as shown in Table \ref{Table_1.2}. \vspace{0.25 cm}

\begin{table*}
\centering
\caption{Coefficients of determination $R^{2}$ and mean absolute error $MAE_{1}$ for the simulations made of restricted $LF(\alpha)$ and restricted $GLF(\alpha)$ at different time instants. The status column corresponds to the literal indicated in Figure \ref{Fig_1.2} and Figure \ref{Fig_1.3}.}
\label{Table_1.2}
\begin{tabular}{@{}ccccccc@{}}
\toprule
Status & $t_{LF}$ & $R_{LF}^{2}(\%)$ & $MAE_{1}^{(LF)}(\times10^{-5})$ & $t_{GLF}$ & $R_{GLF}^{2}(\%)$ & $MAE_{1}^{(GLF)}(\times10^{-7})$ \\ \midrule
A & $6.25$  & $99.94$ & $7.19$ & $5.0$  & $98.89$ & $233.0$\\
B & $12.50$ & $99.99$ & $3.96$ & $10.0$ & $97.99$ & $79.50$\\
C & $18.75$ & $99.98$ & $4.19$ & $15.0$ & $96.65$ & $59.20$\\
D & $25.00$ & $99.97$ & $4.49$ & $20.0$ & $94.59$ & $29.80$\\
E & $31.25$ & $99.99$ & $3.26$ & $25.0$ & $97.26$ & $9.30$\\
F & $37.50$ & $99.99$ & $2.77$ & $30.0$ & $94.52$ & $10.01$\\
G & $43.75$ & $99.99$ & $2.81$ & $35.0$ & $93.61$ & $6.90$\\
H & $50.00$ & $99.99$ & $2.48$ & $40.0$ & $94.57$ & $3.50$\\
\bottomrule
\end{tabular}
\end{table*}

\noindent Therefore, for the four stochastic processes addressed, restricted $BM$, restricted $GBM$, restricted $LF(\alpha)$ and restricted $GLF(\alpha)$, the validity of the analytical solutions of Eq. \eqref{Eq. Entropy Production 10}, \eqref{Eq. Entropy Production 11}, \eqref{Eq. Entropy Production 12}, and \eqref{Eq. Entropy Production 13}, respectively, are verified, which is supported by the obtained $R^{2}$ not less than $93\%$ in all cases. Likewise, it is worth highlighting these four stochastic processes. However, they are a modification of some already known stochastic processes such as the standard $BM$, the standard $GBM$, and the standard $LF(\alpha)$, allow us to introduce with a certain level of arbitrariness a new class of stochastic processes with thresholds. Indeed, it is emphasized that the $GLF(\alpha)$ has not been studied before in the literature and in which Itô's lemma was not necessary to calculate its solution as it is done for the $GBM$. However, the question remains about the asymptotic behavior of this type of stochastic process, that is, the system's behavior for the long-time regime or in the regime of what is known as the stationary limit of the probability distribution. To this end, we conclude with the following subsection that shows the entropy production rate for two of the four stochastic processes addressed. \vspace{0.07 cm}

\subsection{\label{Entropy production}Entropy production in some stochastic processes with threshold}

\noindent In this subsection, the entropy production rate of the restricted $BM$ and the restricted $GBM$ is shown. From this production rate, the usual stationary limit of these two stochastic processes is defined as a quasi-equilibrium state. To do this, we start by defining the Renyi entropy as a measure of the information of the stochastic process, which generalizes the Shannon entropy through the following definition in the continuum
\begin{align}
    \mathbb{H}_{q}\left[\Psi,t\right]&= \lim_{s\to q}{\left[b_{s}+\frac{a_{s}}{1-s}\ln{\left(\int_{-\infty}^{\infty}\Psi^{s}(x,t)dx\right)}\right]}\notag\\
    \label{Eq. Entropy Production 16}
    &=\lim_{s\to q}{\left[b_{s}+\frac{a_{s}}{1-s}\ln{\mathbb{E}\left[\Psi^{s-1}(x,t)\right]}\right]},
\end{align}
\noindent where $a_{s}$ is the amplitude of the Renyi entropy that makes said quantity extensive, and $b_{s}$ is the gauge of the Renyi entropy used to define its zero point. Note that the expression \eqref{Eq. Entropy Production 16} is a continuum generalization of the Renyi entropy used to measure fractal exponents in a discrete system \cite{Jiang2019, Abril202406}. Then, $q=0$, $q=1$, and $q=2$ correspond to the usual fractal dimension \cite{Mandelbrot1967}, the information dimension calculated with Shannon entropy \cite{Renyi1959}, and the correlation dimension \cite{Grassberger1983}, respectively. Furthermore, since the Shannon entropy is an extensive quantity, one must have that $a_{1}=1$, and in such case, the Renyi entropy coincides with the usual definition of statistical mechanics and information theory to quantify the uncertainty or randomness in the state of the system since (see Appendix \ref{Appendix D})
\begin{equation*}
    \mathbb{H}_{1}\left[\Psi,t\right]=-\int_{-\infty}^{\infty}\Psi(x,t)\ln{\Psi(x,t)}dx.
\end{equation*}
\noindent Besides, the integral within the logarithm in Renyi entropy is performed over an arbitrary power of the probability density function of the stochastic process $X(t)$. This justifies our focus on restricted $BM$ and restricted $GBM$, as there is no analytical representation for the probability density function of a $LF(\alpha)$. It is also worth mentioning that the probability density function of the $BM$ satisfies $\mathfrak{f}_{BM}\left(z;0,\sqrt{\tau}\right)=\mathfrak{f}_{BM}\left(z/\sqrt{\tau};0,1\right)$, for all $z\in\mathbb{R}$, and for all $\tau>0$, which implies that the notation $\mathfrak{f}_{BM}\left(z;0,\sqrt{\tau}\right)\equiv\mathfrak{f}_{BM}\left(z/\sqrt{\tau}\right)$ is adopted from now on (the same for its survival function). \vspace{0.25 cm}

\noindent Now, by replacing the expressions \eqref{Eq. Entropy Production 10} and \eqref{Eq. Entropy Production 13} in the definition of Renyi entropy (see Eq. \eqref{Eq. Entropy Production 16}), it is obtained that the Renyi entropies of the restricted $BM$ and the restricted $GBM$ are
\begin{widetext}
\begin{align}
    \label{Eq. Entropy Production 17}
    \mathbb{H}_{q}\left[\Psi_{BM},t\right]&= h_{q}+a_{q}\ln{\sqrt{2\pi\sigma_{0}^{2}\tau}}+\frac{a_{q}}{q-1}\left[q\ln{\mathcal{F}_{BM}\left(\frac{z_{V}}{\sqrt{\tau}}\right)}-\ln{\mathcal{F}_{BM}\left(z_{V}\sqrt{\frac{q}{\tau}}\right)}\right],\\
    \label{Eq. Entropy Production 18}
    \mathbb{H}_{q}\left[\Psi_{GBM},t\right]&= g_{q}+ c_{q}\left[\ln{\sqrt{2\pi\sigma_{0}^{2}x_{0}^{2}\tau}}+\mu_{0}\tau+\frac{\sigma_{0}^{2}\tau}{2q}(1-q)\right]+\frac{c_{q}}{q-1}\left[q\ln{\mathcal{F}_{BM}\left(\frac{w_{V}}{\sqrt{\tau}}\right)}-\ln{\mathcal{F}_{BM}\left(w_{1}\sqrt{\frac{q}{\tau}}\right)}\right],
\end{align}
\end{widetext}
\noindent respectively, where $a_{q}$ is the amplitude of the Renyi entropy of the restricted $BM$, $b_{q}$ is the gauge of the Renyi entropy of the restricted $BM$, $c_{q}$ is the amplitude of the Renyi entropy of the restricted $GBM$, $d_{q}$ is the gauge of the Renyi entropy of the restricted $GBM$. Additionally, $h_{q}=b_{q}+\frac{a_{q}\ln{q}}{2(q-1)}$, $g_{q}=d_{q}+\frac{c_{q}\ln{q}}{2(q-1)}$, and $w_{1}=w_{V}-\frac{\sigma_{0}\tau}{q}(1-q)$. Thence, it is observed that the Renyi entropy for the restricted $BM$ or the restricted $GBM$ has a similar structure where the first factor $h_{q}$ or $g_{q}$, respectively, do not explicitly depend on time. \vspace{0.25 cm}

\noindent Due to the natural connection between statistical mechanics entropy and information entropy in information theory, the principle of maximum entropy suggests that the probability distribution best representing the current state of knowledge about a system is the one with the highest entropy or minimum entropy production. In the steady state of a stochastic process, which typically corresponds to the equilibrium limit in statistical mechanics (equilibrium steady state), the probability distribution should maximize the number of possible configurations in the system, i.e., it should maximize entropy. Also, in practice, the stationary distribution is often defined as the point where the expected value of the Feynman-Kac functional $\Psi(x,t)$ no longer exhibits explicit time variations, reducing the $FP-GIDP(\eta,\gamma)$ to an ordinary differential equation whose solution determines a steady state of the system. Thus, the stationary distribution is the one that maximizes entropy over time or minimizes the entropy production rate. It is important to remember that entropy production does not necessarily disappear in a steady state (non-equilibrium steady state). Consequently, by estimating the temporal variation of the Renyi entropy, the steady limit of a GIDP is estimated even if this temporal variation does not become completely zero. \vspace{0.25 cm}

\noindent Then, taking the time derivative of the Renyi entropy and taking the limit $q\to1$, the Shannon entropy production rate for the restricted $BM$ and restricted $GBM$ are
\begin{widetext}
\begin{align}
    \label{Eq. Entropy Production 19}
    \frac{d\mathbb{H}_{1}}{dt}\left[\Psi_{BM},t\right]&= \frac{1}{2\tau}+\frac{z_{1}\;\mathfrak{f}_{BM}\left(\frac{z_{V}}{\sqrt{\tau}}\right)}{4\tau^{3/2}\mathcal{F}_{BM}\left(\frac{z_{V}}{\sqrt{\tau}}\right)}-\frac{z_{1}z_{V}}{4}\left(\frac{\mathfrak{f}_{BM}\left(\frac{z_{V}}{\sqrt{\tau}}\right)}{\tau\mathcal{F}_{BM}\left(\frac{z_{V}}{\sqrt{\tau}}\right)}\right)^{2}+\frac{z_{V}\;\mathfrak{f}_{BM}\left(\frac{z_{V}}{\sqrt{\tau}}\right)}{4\tau^{5/2}\mathcal{F}_{BM}\left(\frac{z_{V}}{\sqrt{\tau}}\right)}\left(z_{1}z_{V}-\tau\right),\\
    \label{Eq. Entropy Production 20}
    \frac{d\mathbb{H}_{1}}{dt}\left[\Psi_{GBM},t\right]&= \frac{1}{2\tau}+\mu_{0}+\frac{w_{4}\;\mathfrak{f}_{BM}\left(\frac{w_{V}}{\sqrt{\tau}}\right)}{4\tau^{3/2}\mathcal{F}_{BM}\left(\frac{w_{V}}{\sqrt{\tau}}\right)}-\frac{w_{2}w_{3}}{4}\left(\frac{\mathfrak{f}_{BM}\left(\frac{w_{V}}{\sqrt{\tau}}\right)}{\tau\mathcal{F}_{BM}\left(\frac{w_{V}}{\sqrt{\tau}}\right)}\right)^{2}+\frac{w_{3}\;\mathfrak{f}_{BM}\left(\frac{w_{V}}{\sqrt{\tau}}\right)}{4\tau^{5/2}\mathcal{F}_{BM}\left(\frac{w_{V}}{\sqrt{\tau}}\right)}\left(w_{2}w_{V}-\tau\right),
\end{align}
\end{widetext}
\noindent respectively, where $\sigma_{0}z_{1}=x_{V}-x_{0}+\mu_{0}\tau$, $\sigma_{0}w_{2}=\ln{(x_{V})}-\ln{(x_{0})}+\mu_{0}\tau$, $w_{3}=w_{V}+2\sigma_{0}\tau$, and $w_{4}=w_{2}+2\sigma_{0}\tau$. Note that in the case of the standard $BM$ or the standard $GBM$, where the threshold value $x_{V}$ is removed, the Shannon entropy takes the form $\mathfrak{h}_{0}+\ln{\sqrt{\tau}}+\mu_{0}\tau$, where $\mathfrak{h}_{0}$ is a time-independent constant and the linear term $\mu_{0}\tau$ vanishes for the standard $BM$. Hence, the Shannon entropy production rate is of the form $\tau^{-1}/2+\mu_{0}$, where the term $\mu_{0}$ vanishes for the standard $BM$. \vspace{0.25 cm}

\begin{figure*}
    \centering
    \includegraphics[scale=0.395]{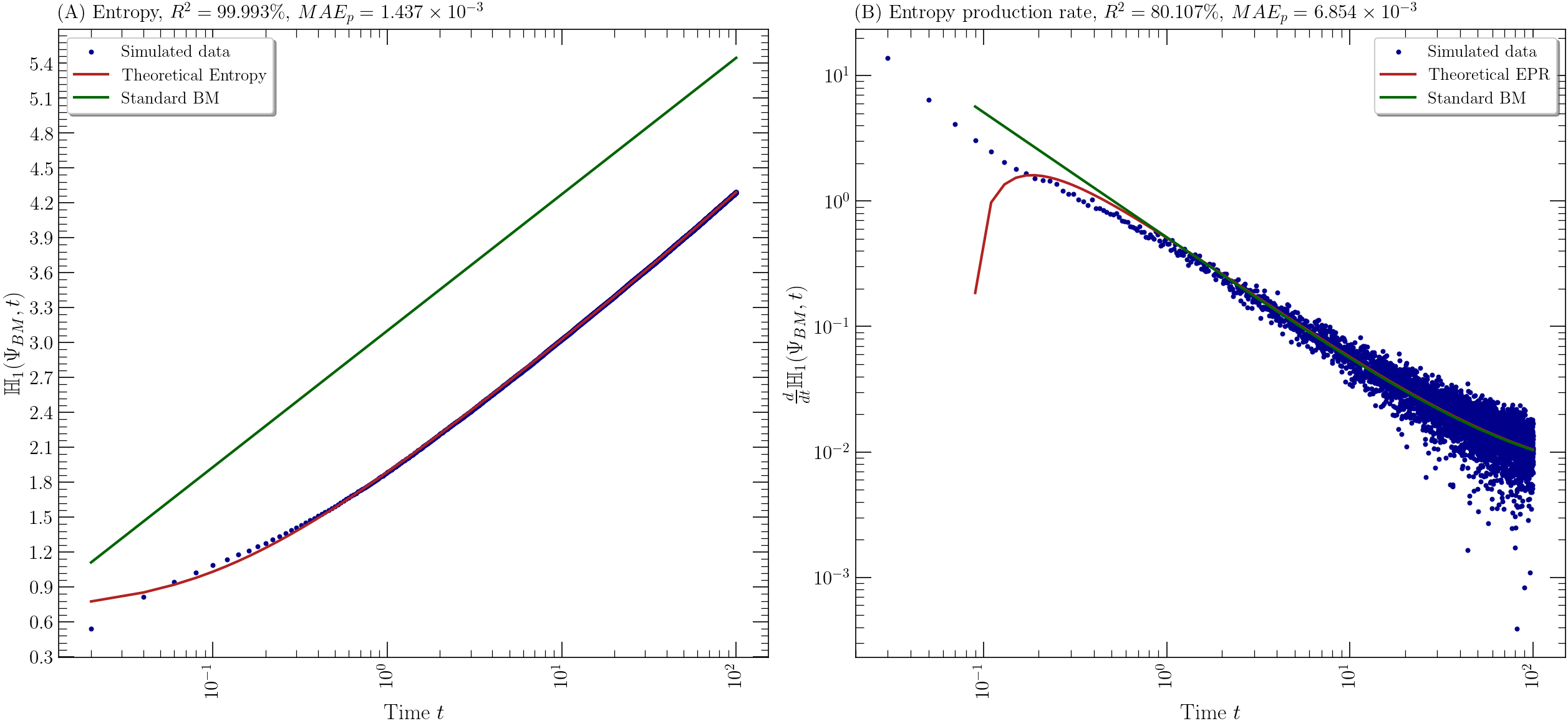}
    \caption{Entropy analysis for restricted Brownian Motion with parameters $x_{0}=2$, $t_{0}=0$, $\mu=1\times10^{-1}$, $\sigma=3$, and $x_{V}=1$, using $N_{s}=4\times10^{4}$ trajectories, $N_{t}=5\times10^{3}$ time steps, $N_{b}=2\times10^{2}$ bins, and a final time of $t_{f}=1\times10^{2}$. (A) Temporal evolution of the Shannon entropy $\mathbb{H}_{1}(\Psi_{BM},t)$. (B) Shannon entropy production rate $\frac{d}{dt}\mathbb{H}_{1}(\Psi_{BM},t)$. In both cases, the red solid line corresponds to the theoretical fits Eq. \eqref{Eq. Entropy Production 17} and Eq. \eqref{Eq. Entropy Production 19}, while the green solid line corresponds to the standard BM case (no threshold) and the points correspond to the simulated data.}
    \label{Fig_1.4}
\end{figure*}

\noindent Figure \ref{Fig_1.4} illustrates the temporal evolution of the Shannon entropy (panel A) and Shannon entropy production rate (panel B) for the restricted $BM$ using the same parameters as Figure \ref{Fig_1.0}. In both cases, the coefficient of determination ($R^{2}$) and mean absolute error ($MAE_{1}$) were calculated, demonstrating a precise fit between the theoretical expressions \eqref{Eq. Entropy Production 17} and \eqref{Eq. Entropy Production 19} with the simulated data. For Shannon entropy, $R^{2}$ is $99.992\%$ and $MAE_{1}$ is $2.026\times10^{-3}$, while for the entropy production rate, $R^{2}$ is $80.063\%$ and $MAE_{1}$ is $6.925\times10^{-3}$. The Shannon entropy gauge $b_{1}$ was determined as the initial entropy value at the first simulation step ($\tau=\frac{t_{f}-t_{0}}{N_{t}}$). Finally, note that when compared to the standard $BM$, a significant difference is observed in the Shannon entropy even though the logarithmic behavior in time is clearly observed thanks to the logarithmic scale in panel A. However, the Shannon entropy production rate behaves similarly in both cases except for a slight deviation around $t\approx0.1$ due to high initial variability and the volatile behavior of the $\mathfrak{f}_{BM}/\mathcal{F}_{BM}$ function, though this does not undermine the validity of the Eq. \eqref{Eq. Entropy Production 19} for restricted $BM$. Note that this is largely because the second and fourth summands of Eq. \eqref{Eq. Entropy Production 19} depends on an odd number of combinations of $z_{V}$ and $z_{1}$, causing the sign of the entropy production rate to alternate depending on the value of the stochastic drift $\mu_{0}$, since $z_{V}<z_{1}<0$ if $x_{V}+\mu_{0}\tau<x_{0}$, which is satisfied for the parameters selected around $\tau\approx0.1$. \vspace{0.25 cm}

\begin{figure*}
    \centering
    \includegraphics[scale=0.395]{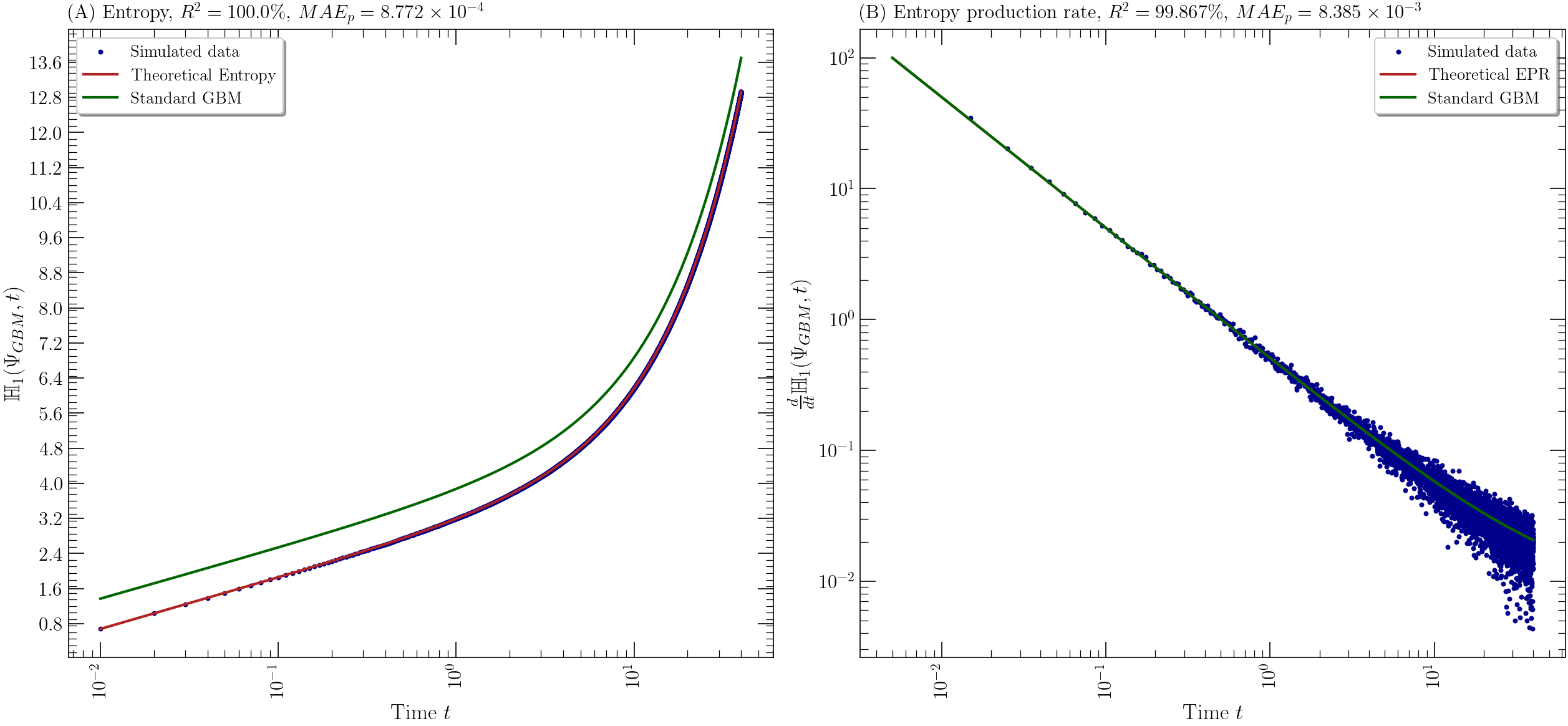}
    \caption{Entropy analysis for restricted geometric Brownian Motion with parameters $x_{0}=6\times10^{1}$, $t_{0}=0$, $\mu=2.05\times10^{-1}$, $\sigma=8\times10^{-2}$, and $x_{V}=1$, using $N_{s}=1\times10^{5}$ trajectories, $N_{t}=4\times10^{3}$ time steps, $N_{b}=4\times10^{2}$ bins, and a final time of $t_{f}=4\times10^{1}$. (A) Temporal evolution of the Shannon entropy $\mathbb{H}_{1}(\Psi_{GBM},t)$. (B) Shannon entropy production rate $\frac{d}{dt}\mathbb{H}_{1}(\Psi_{GBM},t)$. In both cases, the red solid line corresponds to the theoretical fits Eq. \eqref{Eq. Entropy Production 18} and Eq. \eqref{Eq. Entropy Production 20}, while the green solid line corresponds to the standard BM case (no threshold) and the points correspond to the simulated data.}
    \label{Fig_1.5}
\end{figure*}

\noindent Now, Figure \ref{Fig_1.5} illustrates the temporal evolution of the Shannon entropy (panel A) and Shannon entropy production rate (panel B) for the restricted $GBM$ using the same parameters as Figure \ref{Fig_1.1}. In both cases, the coefficient of determination ($R^{2}$) and the mean absolute error ($MAE_{1}$) are recalculated, demonstrating an accurate fit between the theoretical expressions \eqref{Eq. Entropy Production 18} and \eqref{Eq. Entropy Production 20} with the simulated data. For Shannon entropy, $R^{2}$ is $100.00\%$ and $MAE_{1}$ is $8.772\times10^{-4}$, while for the entropy production rate, $R^{2}$ is $99.867 \%$ and $MAE_{1}$ is $8.385\times10^{-3}$. The Shannon entropy gauge $b_{1}$ was determined as the initial entropy value at the first simulation step ($\tau=\frac{t_{f}-t_{0}}{N_{t}}$). Therefore, an even higher precision is observed than for the restricted $BM$. Also, the difference with the Shannon entropy of a standard $BM$ is observed while the Shannon entropy production rate behaves similarly in both cases with a dominant term of the form $\tau^{-1}$. \vspace{0.25 cm}

\noindent Hence, for both restricted $BM$ and restricted $GBM$, the validity of Eq. \eqref{Eq. Entropy Production 17} and Eq. \eqref{Eq. Entropy Production 19} for Shannon entropy is confirmed by taking the limit $q\to1$ on the Renyi entropy. A significant increase in Shannon entropy is observed, indicating that a state of maximum entropy is never reached for these stochastic processes. According to the second law of thermodynamics, since total entropy changes are non-decreasing, the entropy production rate remains non-negative for any time increment $\tau>0$, further verifying the accuracy of Eq. \eqref{Eq. Entropy Production 18} and Eq. \eqref{Eq. Entropy Production 20}. Then, the steady state is defined as the point where the total entropy production rate of the system is zero, consistent with the condition where the Feynman-Kac functional shows no explicit time variation. Although, it is evident that the steady states of the restricted $BM$ and $GBM$ are, in fact, quasi-steady states, as the entropy production rate never reaches zero, remaining on the order of $\mathcal{O}\left(\tau^{-1}\right)$. \vspace{0.07 cm}

\section{\label{Conclusions}Conclusions}

\noindent The multiplicative stochastic path integral is constructed by the Parisi-Sourlas method \cite{Parisi1979, Ovchinnikov2016}. Thus, by extending the SDE of an Itô diffusive process with arbitrary noise (GIDP), one arrives at a general form in which the transition probability between two states is expressed as a usual path integral, i.e. as the integral of an action (see Eq. \eqref{Eq. STS path integral 6}). In fact, from the action, it is observed that the OM functional is a Lagrangian description of an Itô diffusive process, while the MSRJD functional is the Hamiltonian description of an Itô diffusive process (see Eq. \eqref{Eq. STS path integral 10} and Eq. \eqref{Eq. STS path integral 11}, respectively). Likewise, the direct effect of the cumulant-generating function on the action is seen as the analog of the kinetic energy of the system and as the weighting factor that configures the type of trajectories that the system follows in the phase space defined by the stochastic variable $X(t)$ and its canonically conjugate momentum $ip(t)$ (response field). It is also concluded that in the case of a totally random system ($\gamma=0$, $\mu\equiv0$, $\sigma\equiv1$), and scale invariant in the CGF of the noise $\eta(t)$, the transition probability has its solution in terms of the initial parameters of the noise probability density function $\lambda_{1},...,\lambda_{L}$,  and some critical exponents $\nu_{1},...,\nu_{L}$ (see Eq. \eqref{Eq. STS path integral 14}). \vspace{0.25 cm}

\noindent Besides, in section \ref{FP-GIDP} the Fokker-Planck equation of a generalized Itô diffusive process for noise $\eta(t)$ and parameterization $\gamma\in[0,1]$ is calculated ($FP-GIDP(\eta,\gamma)$). For this purpose, the expectation value of the Feynman-Kac functional ($\Psi(x,t)$) is defined as that physical quantity where the effect of the absorptive rate to the action of the stochastic path integral for a $GIDP$ is introduced (see Eq. \eqref{Eq. Fokker-Planck 3}), and a source term $f(x,t)$ that interacts with the absorptive rate since the system starts to evolve. Thus, Eq. \eqref{Eq. Fokker-Planck 7} represents the obtained Fokker-Planck equation where there is an explicit dependence on the type of stochastic calculus used unless the stochastic drift ($\mu\left[X(t),t\right]$) is proportional to the diffusive coefficient ($\sigma\left[X(t),t\right]$). Then, considering the case of scale invariance on the noise distribution $\eta(t)$ and homogeneous and constant coefficients, the $FP-GIDP(\eta,\gamma)$ is solved by showing that $\Psi(x,t)$ can also be interpreted as a transition probability that decays more rapidly due to the effect of a dissipative factor that depends on the height of the potential barrier $V_{0}$. Also, the $FP-GIDP(\eta,\gamma)$ is solved when the stochastic drift is proportional to the diffusive coefficient by showing that in this case, we return to an equation with constant and homogeneous coefficients. It is worth noting that Eq. \eqref{Eq. Fokker-Planck 7} can also accommodate cases involving memory kernels when the stochastic drift is expressed as an integral operator. However, a detailed exploration of this extension is left for future work. \vspace{0.25 cm}

\noindent Finally, in section \ref{Stochastic Processes application}, by introducing an infinite potential barrier around the value $x_{V}$ (absorbing boundary condition), we define the stochastic processes restricted Brownian motion (see Eq. \eqref{Eq. Entropy Production 8}), restricted geometric Brownian motion (see Eq. \eqref{Eq. Entropy Production 9}), Levy $\alpha$-stable flight (see Eq. \eqref{Eq. Entropy Production 10}), and geometric Levy $\alpha$-stable flight (see Eq. \eqref{Eq. Entropy Production 11}). In all cases, the expectation value of the Feynman-Kac functional $\Psi(x,t)$ corresponds to the transition probability of the system such that $\Psi(x,t)=0$, for all $x<x_{V}$. Also, the first three types of stochastic processes are modifications of some stochastic processes already known in the literature. Still, the last stochastic process, the geometric Levy $\alpha$-stable flight, is a new type of stochastic process whose solution was found without the need of Itô's lemma, which is not valid for an $\alpha$-stable distribution in general. Thus, a quite precise and accurate correspondence of the analytical solution of $\Psi(x,t)$ at different instants of time is observed in section \ref{Evolution PDF}. At last, the Shannon entropy and the Shannon entropy production rate of the restricted $BM$ and the restricted $GBM$ are calculated because they have analytical probability density functions. Thus, there is again a precise and accurate correspondence of the analytical solution with the simulated data. In fact, from these results, it is found that the restricted $BM$ and the restricted $GBM$ do not have a steady state for their probability density function but a quasi-steady state since their entropy production rate never vanishes. \vspace{0.25 cm}

\noindent As a final comment, it is noted that the Fokker-Planck equation of a generalized Itô diffusive process unifies many of the results found in the literature for Langevin equations. All this is based on the action of the multiplicative stochastic path integral which depends explicitly on the parameterization $\gamma\in[0,1]$ when the stochastic drift is not proportional to the diffusive coefficient. Thus, our findings have implications for various fields, including statistical physics, econophysics, and biological modeling, where multiplicative noise and non-equilibrium effects play a crucial role. Likewise, the results presented in this work were limited to "geometric" type processes, but can easily be extended to other types of stochastic processes where the stochastic drift and the diffusive coefficient present much more complex forms but proportional to each other to be able to find their analytical solution more simply. Furthermore, as a natural extension of this work that goes beyond the scope of this paper, the generalization of the stochastic path integral formalism to multiple variables and higher-dimensional systems is proposed for future research. This would allow the study of coupled stochastic processes, interactions between degrees of freedom, and correlation effects in non-equilibrium dynamics, offering deeper insights into complex stochastic systems. \vspace{0.25 cm}


\begin{acknowledgments}
\noindent We wish to acknowledge the support of the Economics and Business Department of Universidad de Almería, and the Physics Department of Universidad Nacional de Colombia. \vspace{0.25 cm}

\noindent J. E. Trinidad-Segovia and M. A. Sánchez-Granero are supported by grant PID2021-127836NB-I00 (Ministerio Español de Ciencia e Innovación and FEDER). \vspace{0.25 cm}

\noindent F. S. A. and C. J. Q.  conceived the original idea and developed the theoretical formalism; F. S. A. performed the analytic calculations and the numerical simulations. C. J. Q., J. E. T., and M. A. S. supervised the findings of this work and administered the project. All authors discussed the results and contributed to the final manuscript. \vspace{0.25 cm}
\end{acknowledgments}


\bibliography{apssamp}

\vspace{0.07 cm}

\appendix

\section{\label{Appendix A}Determinant of the Jacobian of the stochastic path integral}

\noindent To calculate $\mathcal{J}\left[X(t),\eta\left[X(t)\right]\right]$ in Eq. \eqref{Eq. STS path integral 5} the following result is used which establishes that for a diagonalizable matrix $\mathcal{B}$, of size $N\in\mathbb{Z}^{+}$, we have that
\begin{align}
    \det{\left(I+\varepsilon\mathcal{B}\right)} &= 1+\varepsilon\text{Tr}\left[\mathcal{B}\right]\notag\\
    \label{Eq. STS Jacobian 1}
    &\hspace*{0.4 cm}+\frac{\varepsilon^{2}}{2}\left[\text{Tr}^{2}\left[\mathcal{B}\right]-\text{Tr}\left[\mathcal{B}^{2}\right]\right]+\mathcal{O}\left(\varepsilon^{3}\right).
\end{align}
\noindent The proof of Eq. \eqref{Eq. STS Jacobian 1} lies in assuming that $\det{\left(I+\varepsilon\mathcal{B}\right)}=\sum_{m=0}^{M}\varepsilon^{m}f_{m}\left(\mathcal{B}\right)$, with $M=2\in\mathbb{N}$, and remembering that $f_{0}\left[\mathcal{B}\right]=1$, $f_{1}\left[\mathcal{B}\right]=\text{Tr}\left[\mathcal{B}\right]$, $\det{\left(\exp{\left(\varepsilon\mathcal{B}\right)}\right)}=\exp{\left(\varepsilon\text{Tr}\left[\mathcal{B}\right]\right)}$, since these equalities imply that
\begin{widetext}
\begin{align}
    \label{Eq. STS Jacobian 2}
    \det{\left[I+\varepsilon\left(\mathcal{B} +\frac{\varepsilon}{2}\mathcal{B}^{2}+\cdots\right)\right]} &= \det{\left(\exp{\left(\varepsilon\mathcal{B}\right)}\right)}=\exp{\left(\varepsilon\text{Tr}\left[\mathcal{B}\right]\right)}= 1+\varepsilon\text{Tr}\left[\mathcal{B}\right]+\frac{\varepsilon^{2}}{2}\text{Tr}^{2}\left[\mathcal{B}\right]+\mathcal{O}\left(\varepsilon^{3}\right)
\end{align}
\begin{align}
    \det{\left[I+\varepsilon\left(\mathcal{B}+ \frac{\varepsilon}{2}\mathcal{B}^{2}+\cdots\right)\right]} &= \sum_{m=0}^{M}\varepsilon^{m}f_{m}\left[\mathcal{B}+\frac{\varepsilon}{2}\mathcal{B}^{2}+\cdots\right]\notag\\
    &= 1+ \varepsilon\text{Tr}\left[\mathcal{B}+\frac{\varepsilon}{2}\mathcal{B}^{2}+\cdots\right]+\varepsilon^{2}f_{2}\left[\mathcal{B}+\frac{\varepsilon}{2}\mathcal{B}^{2}+\cdots\right]+\mathcal{O}\left(\varepsilon^{3}\right)\notag\\
    \label{Eq. STS Jacobian 3}
    &=1+ \varepsilon\text{Tr}\left[\mathcal{B}\right]+\frac{\varepsilon^{2}}{2}\text{Tr}\left[\mathcal{B}^{2}\right] +\varepsilon^{2}f_{2}\left[\mathcal{B}+\frac{\varepsilon}{2}\mathcal{B}^{2}+\cdots\right]+\mathcal{O}\left(\varepsilon^{3}\right)
\end{align}    
\end{widetext}
\noindent Then, by subtracting Eq. \eqref{Eq. STS Jacobian 2} and Eq. \eqref{Eq. STS Jacobian 3} we obtain that
\begin{align}
    f_{2}\left[\mathcal{B}\right] &=\lim_{\varepsilon\to0}{f_{2}\left[\mathcal{B}+\frac{\varepsilon}{2}\mathcal{B}^{2}+\cdots\right]}\notag\\
    \label{Eq. STS Jacobian 4}
    &=\frac{1}{2}\left[\text{Tr}^{2}\left[\mathcal{B}\right]-\text{Tr}\left[\mathcal{B}^{2}\right]\right].
\end{align}
\noindent Thus, from Eq. \eqref{Eq. STS Jacobian 1} it follows that by taking $\varepsilon=\Delta t$ and $\mathcal{B}$ as a function dependent on time $t$, and stochastic variable $X(t)$, with $N\to\infty$, the following expression is satisfied
\begin{align}
    \mathcal{J}&\sim\lim_{N\to\infty}{\det{\left(I-\Delta t\;\mathcal{B}\left[X(t),t\right]\right)}}\notag\\
    &\approx\lim_{N\to\infty}{\left[1-\text{Tr}\left[\mathcal{B}\right]\Delta t+\mathcal{O}\left(\left(\Delta t\right)^{2}\right)\right]}\notag\\
    \label{Eq. STS Jacobian 5}
    &=\exp{\left\{-\int_{t_{0}}^{t_{f}}\text{Tr}\left[\mathcal{B}\left[X(t),t\right]\right]dt\right\}}.
\end{align}
\noindent Also, from the definition of the functional derivative and the definition of the generalized Itô diffusive process (see Eq. \eqref{Eq. Cumulant Generating Function 2}), we have that the noise $\eta(t)$ is expressed in the discrete in terms of stochastic variable $X(t)$ as
\begin{equation}
    \label{Eq. STS Jacobian 6}
    \eta_{j+1}\Delta t=\frac{1}{\sigma_{j+1}}\left[\Delta X-\gamma\mu_{j+1}\Delta t-(1-\gamma)\mu_{j}\Delta t\right]
\end{equation}
\noindent where $X_{j}=X(t_{j})$, $\Delta X=X_{j+1}-X_{j}$, $\eta_{j}=\eta(t_{j})$,  $\mu_{j}=\mu\left[X_{j},t_{j}\right]$, $\sigma_{j}=\sigma\left[X_{j},t_{j}\right]$, for all $j\in\{0,1,...,N\}$, and $\left\{t_{0},t_{1},...,t_{N}\right\}$ a partition of $\left[t_{0},t_{f}\right]$. Thus, $t_{0}<t_{1}<\cdots<t_{N}=t_{f}$, and $\Delta t=\max{\left\{t_{j+1}-t_{j}\left|0\leq j\leq N-1\right.\right\}}$. Likewise, $\gamma\in\left[0,1\right]$ represents a parameterization of stochastic calculus such that $\gamma=0$ corresponds to the Itô calculus \cite{Ito1950}, $\gamma=1/2$ corresponds to the Fisk-Stratonovich calculus \cite{Stratonovich1966}, and $\gamma=1$ corresponds to the Hänggi-Klimontovich calculus \cite{Hanggi1982}. \vspace{0.25 cm}

\noindent Now, note that $\eta_{j}\Delta t$ represents the differential noise increments since if we assume that there exists a stochastic variable $\beta$ such that $\frac{d\beta}{dt}=\eta(t)$, then $\eta_{j}\Delta t=\beta_{j+1}-\beta_{j}$, with $\beta_{j}=\beta(t_{j})$. Thence, taking the right-hand side of Eq. \eqref{Eq. STS Jacobian 6} directly, we have that $\mathcal{J}\left[X(t),\eta\left[X(t)\right]\right]$ is
\begin{widetext}
\begin{align}
    \mathcal{J}\left[X(t),\eta\left[X(t)\right]\right] &\equiv\lim_{N\to\infty}{\mathcal{J}\left[X_{0},X_{1},...,X_{N},\eta_{0},\eta_{1},...,\eta_{N}\right]}=\lim_{N\to\infty}{\left|\det{\left(\frac{\partial\eta_{i}}{\partial X_{k}}\right)}\right|}\notag\\
    &=\lim_{N\to\infty}{\left|\det{\left[\frac{\partial}{\partial X_{k}}\left(\frac{X_{i+1}-X_{i}-\gamma\mu_{i+1}\Delta t-\left(1-\gamma\right)\mu_{i}\Delta t}{\sigma_{i+1}}\right)\right]}\right|}\notag\\
    &=\lim_{N\to\infty}{\left|\det{\left[\frac{1}{\sigma_{i+1}}\left(\delta_{i+1,k}-\delta_{i,k}-\gamma\delta_{i+1,k}\frac{\partial \mu_{i+1}}{\partial X_{i+1}}\Delta t-(1-\gamma)\delta_{i,k}\frac{\partial \mu_{i}}{\partial X_{i}}\Delta t\right)\right.}\right.}\notag\\
    &\hspace*{1.0 cm}-\left.\left.\frac{1}{\sigma_{i+1}^{2}}\frac{\partial\sigma_{i+1}}{\partial X_{i+1}}\delta_{i+1,k}\left(X_{i+1}-X_{i}-\gamma\mu_{i+1}\Delta t-(1-\gamma)\mu_{i}\Delta t\right)\right]\right|\notag\\
    &=\lim_{N\to\infty}{\frac{1}{\prod_{k=0}^{N-1}\sigma_{k}}\prod_{k=0}^{N-1}\left|1-\gamma\frac{\partial\mu_{k}}{\partial X_{k}}\Delta t-\frac{1}{\sigma_{k}}\frac{\partial\sigma_{k}}{\partial X_{k}}\left(X_{k+1}-X_{k}-\gamma\mu_{k}\Delta t\right)\right|}\notag\\
    &=\lim_{N\to\infty}{e^{-\sum_{k=0}^{N-1}\ln{\left(\sigma_{k}\right)\frac{\Delta t}{\Delta t}}}\prod_{k=0}^{N-1}\left|1-\Delta t\left(\gamma\frac{\partial\mu_{k}}{\partial X_{k}}-\gamma\mu_{k}\frac{\partial\ln{\sigma_{k}}}{\partial X_{k}}-\frac{X_{k+1}-X_{k}}{\Delta t}\frac{\partial\ln{\sigma_{k}}}{\partial X_{k}}\right)\right|}\notag\\
    \label{Eq. STS Jacobian 7}
    &=\exp{\left\{\int_{t_{0}}^{t_{f}}\left[\gamma\mu\left[X,t\right]\frac{\partial\ln{\sigma\left[X,t\right]}}{\partial X}-\gamma\frac{\partial\mu\left[X,t\right]}{\partial X}-\frac{dX}{dt}\frac{\partial\ln{\sigma\left[X,t\right]}}{\partial X}-\frac{\partial\ln{\sigma\left[X,t\right]}}{\partial t}\right]dt\right\}}.
\end{align}
\end{widetext}
\noindent Finally, note that the last two terms of Eq. \eqref{Eq. STS Jacobian 7} corresponds to a total time derivative since
\begin{align}
    \label{Eq. STS Jacobian 8}
    \frac{d\xi}{dt}&=\frac{\partial\xi}{\partial t}+\frac{\partial\xi}{\partial X}\frac{dX}{dt},
\end{align}
\noindent where $\xi\left[X(t),t\right]=\ln{\sigma\left[X(t),t\right]}$. Thus, this total time derivative would give a constant of integration in Eq. \eqref{Eq. STS Jacobian 7}, but since the transition probability satisfies a normalization condition (see Eq. \eqref{Eq. Cumulant Generating Function 6}), it will be seen that this constant does not affect the dynamics of the process, in a manner analogous to the principle of least action with a total time derivative, which allows us to conclude that $\mathcal{J}\left[X\right]\equiv\mathcal{J}\left[X(t),\eta(t)\right]$ is
\begin{align}
    \mathcal{J}[X]&=\exp{\left\{-\gamma\int_{t_{0}}^{t_{f}}\left[\frac{\partial\mu}{\partial X}-\mu\frac{\partial\ln{\sigma}}{\partial X}\right]dt\right\}}\notag\\
    \label{Eq. STS Jacobian 9}
    &=\exp{\left\{-\gamma\int_{t_{0}}^{t_{f}}\sigma\frac{\partial}{\partial X}\left(\frac{\mu}{\sigma}\right)dt\right\}}.
\end{align}

\section{\label{Appendix B}Additive drift-free propagator}
\noindent The additive drift-free stochastic path integral, i.e., $\sigma\equiv1$ and $\mu\equiv0$, has an analytical expression for the probability transition amplitude defined in Eq. \eqref{Eq. STS path integral 6}. To compute it, one proceeds analogously to Lemma $3.1$ in \cite{Issaka2016}. Specifically, from the definition of functional integration, we have that $\mathcal{P}_{\eta}\left(X_{f},t_{f}|X_{0},t_{0}\right)$ corresponds to
\begin{align}
    \mathcal{P}_{\eta}\left(X_{f},t_{f}|X_{0},t_{0}\right) &= \int_{X_{0}}^{X_{f}}\mathcal{D}X\int\frac{\mathcal{D}p}{2\pi}e^{\mathcal{S}_{\eta}\left[\dot{X}(t),X(t)\right]}\notag\\
    &=\lim_{N\to\infty}{\int_{\mathbb{R}}\frac{dp_{N}}{2\pi}\prod_{n=0}^{N-1}\left[\int_{\mathbb{R}}dX_{n}\int_{\mathbb{R}}\frac{dp_{n}}{2\pi}\right]} \notag\\
    \label{Eq. Drift-free 1}
    &\hspace*{0.4 cm}\times\exp{\left\{\sum_{n=0}^{N-1}\mathcal{S}_{\eta}\left[\dot{X}_{n},X_{n}\right]\right\}},
\end{align}
\noindent where an additional integral over $p(t)$ is taken in the same way as the usual path integral, $\varepsilon=\frac{t_{f}-t_{0}}{N}$ is the infinitesimal time step, $t_{n}=n\varepsilon$, $X_{n}=X(t_{n})$, for all $n\in\{0, 1, 2,\dots, N-1\}$, and $X_{N}=X_{f}$. \vspace{0.25 cm}

\noindent Now, note that the stochastic path integral Lagrangian satisfies the following identity
\begin{align*}
    \mathcal{I}&=\sum_{n=0}^{N-1}\mathcal{S}_{\eta}\left[\dot{X}_{n},X_{n}\right]=\varepsilon\sum_{n=0}^{N-1}\mathcal{L}_{\eta}\left[\dot{X}_{n},X_{n}\right]\\
    &= \sum_{n=0}^{N-1}\left[ip_{n+1}\left(X_{n+1}-X_{n}\right)+\varepsilon\mathcal{K}_{\eta}\left[p_{n+1}\right]\right]+\varepsilon\mathcal{K}_{\eta}\left[p_{0}\right]
\end{align*}

\begin{widetext}
\begin{align}
    \prod_{n=0}^{N-1}\left[\int_{\mathbb{R}^{2}}dX_{n}\right]\;e^{\mathcal{I}}&= \prod_{n=0}^{N-1}\left[\int_{\mathbb{R}^{2}}dX_{n}\frac{dp_{n}}{2\pi}\right]\exp{\left\{i\sum_{n=0}^{N-1}\left[p_{n+1}X_{n+1}-p_{n}X_{n}\right]+\varepsilon\sum_{n=0}^{N}\mathcal{K}_{\eta}\left[p_{n}\right]-i\sum_{n=0}^{N-1}X_{n}\left(p_{n+1}-p_{n}\right)\right\}}\notag\\
    &= \prod_{n=0}^{N-1}\left[\int_{\mathbb{R}^{2}}dX_{n}\frac{dp_{n}}{2\pi}\right]\exp{\left\{ip_{N}X_{N}-ip_{0}X_{0}+\varepsilon\sum_{n=0}^{N}\mathcal{K}_{\eta}\left[p_{n}\right]-i\sum_{n=0}^{N-1}X_{n}\left(p_{n+1}-p_{n}\right)\right\}}\notag\\
    &= e^{ip_{N}X_{f}+\varepsilon\mathcal{K}_{\eta}\left[p_{N}\right]}\prod_{n=0}^{N-1}\int_{\mathbb{R}}\int_{\mathbb{R}}e^{-iX_{n}\left(p_{n+1}-p_{n}\right)}dX_{n}\;e^{\varepsilon\mathcal{K}_{\eta}\left[p_{n}\right]-ip_{0}X_{0}}\;\frac{dp_{n}}{2\pi}\notag\\
    &= e^{ip_{N}X_{f}+\varepsilon\mathcal{K}_{\eta}\left[p_{N}\right]}\prod_{n=0}^{N-1}\int_{\mathbb{R}}\delta(p_{n+1}-p_{n})e^{\varepsilon\mathcal{K}_{\eta}\left[p_{n}\right]-ip_{0}X_{0}}dp_{n}\notag\\
    &= e^{ip_{N}X_{f}+\varepsilon\mathcal{K}_{\eta}\left[p_{N}\right]}\;e^{-ip_{N}X_{0}+\varepsilon(N-1)\mathcal{K}_{\eta}\left[p_{N}\right]}\notag\\
    \label{Eq. Drift-free 2}
    &= e^{ip_{N}\left(X_{f}-X_{0}\right)+(t_{f}-t_{0})\mathcal{K}_{\eta}\left[p_{N}\right]}.
\end{align}
\end{widetext}

\noindent Consequently, by replacing Eq. \eqref{Eq. Drift-free 2} in Eq. \eqref{Eq. Drift-free 1} it is obtained
\begin{align}
    \mathcal{P}_{\eta}\left(X_{f},t_{f}|X_{0},t_{0}\right) &= \lim_{N\to\infty}{\int_{\mathbb{R}}\frac{dp_{N}}{2\pi}e^{ip_{N} \Delta X+\tau\mathcal{K}_{\eta}\left[p_{N}\right]}}\notag\\
    \label{Eq. Drift-free 3}
    &=\int_{-\infty}^{\infty}e^{ip\Delta X+\tau\mathcal{K}_{\eta}(p)}\frac{dp}{2\pi},
\end{align}
\noindent where $\tau=t_{f}-t_{0}$, and $\Delta X=X_{f}-X_{0}$. \vspace{0.07 cm}

\section{\label{Appendix C}Differential increase in the expected value of Feynman-Kac functional}
\noindent Let $\varepsilon\to0^{+}$ and $X(t+\varepsilon)=z$, then the expected value $\Psi(z,t+\varepsilon)\equiv\left\langle\psi\left(z,t+\varepsilon|x,t\right)\right\rangle$ of the Feynman-Kac functional defined in Eq. \eqref{Eq. Fokker-Planck 5} in an infinitesimal time interval is
\begin{widetext}
\begin{align}
    \Psi\left(z,t+\varepsilon\right)&=\int_{-\infty}^{\infty}\mathfrak{m}_{0}(dz;x,t,t+\varepsilon)\psi\left(z,t+\varepsilon|x,t\right)=\int_{-\infty}^{\infty}\mathcal{P}_{\eta}(z,t+\varepsilon|x,t)\psi\left(z,t+\varepsilon|x,t\right)dz\notag\\
    &=\int_{X(t)=x}\int e^{\int_{t}^{t+\varepsilon}\mathcal{L}_{\eta}\left[\dot{X}(\tau),X(\tau),\tau\right]d\tau}\psi\left[X(t+\varepsilon)=z,t+\varepsilon|X(t)=x,t\right]\frac{\mathcal{D}p}{2\pi}\mathcal{D}X\notag\\    
    &\approx\int_{-\infty}^{\infty}\int_{-\infty}^{\infty}e^{ip\varepsilon\left(\frac{z-x}{\varepsilon}\right)-\int_{t}^{t+\varepsilon}\mathcal{H}_{\eta}\left[p(\tau),X(\tau)\right]d\tau}\psi(z,t+\varepsilon|x,t)\frac{dp}{2\pi}dx\notag\\
    &\approx\int_{-\infty}^{\infty}\left(\int_{-\infty}^{\infty}e^{ip(z-x)}\left[1-\varepsilon\mathcal{H}_{\eta}(p,x)\right]\frac{dp}{2\pi}\right)\psi(z,t+\varepsilon|x,t)dx+\mathcal{O}\left(\varepsilon^{2}\right)\notag\\
    &=\int_{-\infty}^{\infty}\int_{-\infty}^{\infty}e^{ip(z-x)}\frac{dp}{2\pi}\left[1-\varepsilon\mathcal{H}_{\eta}\left(x,-i\frac{\partial}{\partial x}\right)\right]\psi(z,t+\varepsilon|x,t)dx+\mathcal{O}\left(\varepsilon^{2}\right)\notag\\
    &=\int_{-\infty}^{\infty}\delta(z-x)\left[1-\varepsilon\mathcal{H}_{\eta}\left(x,-i\frac{\partial}{\partial x}\right)\right]\left[\psi^{*}(z,t|x,t)+\varepsilon\left.\frac{d}{d\varepsilon}\psi(z,t+\varepsilon|x,t)\right|_{\varepsilon=0}\right]dx+\mathcal{O}\left(\varepsilon^{2}\right)\notag\\
    &=\int_{-\infty}^{\infty}\delta(z-x)\left[1-\varepsilon\mathcal{H}_{\eta}\left(x,-i\frac{\partial}{\partial x}\right)\right]\left[\psi^{*}(z,t|x,t)+\varepsilon f(x,t)-\varepsilon\mathcal{V}(x,t)\psi^{*}(z,t|x,t)\right]dx+\mathcal{O}\left(\varepsilon^{2}\right)\notag\\
    \label{Eq. Feynman-Kac 1}
    &=\Psi(z,t)+\varepsilon\left[f(z,t)-\mathcal{V}(z,t)\Psi(z,t)-\mathcal{H}_{\eta}\left[z,-i\frac{\partial}{\partial z}\right]\Psi(z,t)\right]+\mathcal{O}\left(\varepsilon^{2}\right),
\end{align} 
\end{widetext}
\noindent where $\psi^{*}(z,t|x,t)=\lim_{\varepsilon\to0^{+}}\psi(z,t+\varepsilon|x,t)$. Furthermore, in the previous expression's second step, the expected value is defined as a path integral with a single fixed endpoint at $X(t)=x$. Thus, it is worth remembering that in the definition of the functional measures of the Feynman path integral where there are two fixed ends, there always exists an additional integral over the moments, which implies that with a single fixed endpoint, the number of integrals over the generalized coordinates and the canonically conjugate moments is equalized which explains the approximation of the third step in Eq. \eqref{Eq. Feynman-Kac 1}. Also, in the fifth step of Eq. \eqref{Eq. Feynman-Kac 1} the definition of CGF as an analytical function (see Eq. \eqref{Eq. Cumulant Generating Function 5}), and the property of plane waves as eigenfunctions of the canonically conjugate moment, i.e. $pe^{ipx}=-i\frac{\partial}{\partial x}e^{ipx}$, was taken into account. In the seventh step of \eqref{Eq. Feynman-Kac 1}, also the Eq. \eqref{Eq. Fokker-Planck 6} was used. Finally, it is emphasized that the limits containing $\psi^{*}(z,t|x,t)$ must be taken with care since $\lim{\varepsilon\to0^{+}}{\mathcal{P}_{\eta}(z,t+\varepsilon|x,t)}=\delta(z-x)$, implies
\begin{align}
    \Psi(z,t)&=\int_{-\infty}^{\infty}\mathcal{P}_{\eta}(z,t|x,t)\psi(z,t|x,t)dz\notag\\
    \label{Eq. Feynman-Kac 2}
    &=\int_{-\infty}^{\infty}\delta(z-x)\psi^{*}(z,t|x,t).
\end{align}
\noindent Therefore, taking the limit $\varepsilon\to0^{+}$ and dividing by $\varepsilon$, with the first term of Eq. \eqref{Eq. Feynman-Kac 1} we obtain a time derivative and the evolution equation of the Feynman-Kac functional is
\begin{align}
    \label{Eq. Feynman-Kac 3}
    \left[\frac{\partial}{\partial t}+\mathcal{V}(z,t)+\mathcal{H}_{\eta}\left(z,-i\frac{\partial}{\partial z}\right)\right]\Psi(z,t)=f(z,t).
\end{align}

\section{\label{Appendix D}Shannon entropy from Renyi entropy}

\noindent To show how Renyi entropy generalizes Shannon entropy, a more well-known formulation of entropy, we consider the specific case $q=1$ in Eq. \eqref{Eq. Entropy Production 16}. In this case, since $\ln{\mathbb{E}\left[1\right]}=\ln{1}=0$, $a_{1}=1$, it follows that (taking $b_{1}=0$ for simplicity)
\begin{align}
    \mathbb{H}_{1}\left[\Psi,t\right]&=\lim_{s\to 1}{\left[b_{s}+\frac{a_{s}}{1-s}\ln{\mathbb{E}\left[\Psi^{s-1}(x,t)\right]}\right]}\notag\\
    &= b_{1}+a_{1}\lim_{s\to1}{\frac{\ln{\mathbb{E}\left[\Psi^{s-1}(x,t)\right]}}{1-s}}\notag\\
    &= \lim_{s\to1}{\left[-\frac{\frac{d}{ds}\mathbb{E}\left[\Psi^{s-1}(x,t)\right]}{\mathbb{E}\left[\Psi^{s-1}(x,t)\right]}\right]}\notag\\
    &= -\lim_{s\to1}{\frac{\int_{-\infty}^{\infty}dx\frac{\partial}{\partial s}\Psi^{s}(x,t)}{\mathbb{E}\left[\Psi^{s-1}(x,t)\right]}}\notag\\
    &= -\lim_{s\to1}{\frac{\int_{-\infty}^{\infty}\Psi^{s}(x,t)\ln{\Psi(x,t)}dx}{\mathbb{E}\left[\Psi^{s-1}(x,t)\right]}}\notag\\
    &= -\lim_{s\to1}{\frac{\mathbb{E}\left[\Psi^{s-1}(x,t)\ln{\Psi(x,t)}\right]}{\mathbb{E}\left[\Psi^{s-1}(x,t)\right]}}\notag\\
    &=-\mathbb{E}\left[\ln{\Psi(x,t)}\right]\notag\\
    \label{Eq. Shannon Entropy 1}
    &=-\int_{-\infty}^{\infty}\Psi(x,t)\ln{\Psi(x,t)}dx,
\end{align}
\noindent which corresponds to the standard definition of differential entropy, also known as Shannon entropy, in the continuous case \cite{Jaynes1957, Jaynes1957_2, Janssen1981}.

\end{document}